\documentclass[reprint, prd]{revtex4-1}

\usepackage[utf8]{inputenc}
\usepackage{amsmath}
\usepackage{amssymb}
\usepackage{mathrsfs}
\usepackage{times}
\usepackage{hyperref}
\usepackage{natbib}
\usepackage{graphicx}
\usepackage{hyperref}
\usepackage{natbib}

\hypersetup{
  bookmarksnumbered = true,
  bookmarksopen=false,
  pdfborder=0 0 0,         
  pdffitwindow=true,      
  pdfnewwindow=true, 
  colorlinks=true,           
  linkcolor=blue,            
  citecolor=magenta,    
  filecolor=magenta,     
  urlcolor=cyan              
}

\newcommand{\nn}{\nonumber}

\newcommand{\msun}{\ensuremath{M_\odot\,}}
\newcommand{\chimera}{{\sc Chimera }}

\newcommand{\gcc}{\ensuremath{{\mbox{g~cm}}^{-3}}}

\newcommand{\apjl}{Astrophys. J. Lett.}
\newcommand{\mnras}{Mon. Not. R. Astron. Soc.}

\begin{document}

\title{The Gravitational Wave Signal of a Core Collapse Supernova Explosion of a 15 M$_\odot$ Star}

\author{Anthony Mezzacappa$^{1,2}$, Pedro Marronetti$^{3}$, Ryan E. Landfield$^{1,4}$, Eric J. Lentz$^{1,2,4,5}$,   \\
Konstantin N. Yakunin$^{1,2,4}$, Stephen W. Bruenn$^{6}$, W. Raphael Hix$^{1,4}$, O. E. Bronson Messer$^{1,4,7}$, \\
Eirik Endeve$^{1,2,7}$, John M. Blondin$^8$, and J. Austin Harris$^{7}$}

\affiliation{$^1$Department of Physics and Astronomy, University of Tennessee, Knoxville, TN 37996-1200, USA}
\affiliation{$^2$Joint Institute for Computational Sciences, Oak Ridge National Laboratory, P.O. Box 2008, Oak Ridge, TN 37831-6354, USA}
\affiliation{$^3$Physics Division, National Science Foundation, Alexandria, VA 22314 USA}
\affiliation{$^4$Physics Division, Oak Ridge National Laboratory, P.O. Box 2008, Oak Ridge, TN 37831-6354, USA}
\affiliation{$^5$Joint Institute for Nuclear Physics and its Applications, Oak Ridge National Laboratory, P.O. Box 2008, Oak Ridge, TN 37831-6374, USA}
\affiliation{$^6$Department of Physics, Florida Atlantic University, 777 Glades Road, Boca Raton, FL 33431-0991, USA}
\affiliation{$^7$National Center for Computational Sciences, Oak Ridge National Laboratory, P.O. Box 2008, Oak Ridge, TN 37831-6164, USA}
\affiliation{$^8$Department of Physics, North Carolina State University,  Raleigh, NC 27695-8202, USA}

\email{mezz@utk.edu}

\begin{abstract}
We report on the gravitational wave signal computed in the context of a three-dimensional simulation of a core collapse supernova explosion of a 15\msun star. The simulation was performed with our neutrino hydrodynamics code \chimera. We detail the gravitational wave strains as a function of time, for both polarizations, and discuss their physical origins. We also present the corresponding spectral signatures. 
Gravitational wave emission in our model has two key features: low-frequency emission ($<$200 Hz) emanates from the gain layer as a result of neutrino-driven convection and the SASI and high-frequency emission ($>$600 Hz) emanates from the proto-neutron star due to Ledoux convection within it. 
The high-frequency emission dominates the gravitational wave emission in our model and emanates largely from the convective layer itself, not from the convectively stable layer above it, due to convective overshoot. Moreover, the low-frequency emission emanates from the gain layer itself, not from the proto-neutron star, due to 
accretion onto it. We provide evidence of the SASI in our model and demonstrate that the peak of our low-frequency gravitational wave emission spectrum corresponds to it. Given its origin in the gain layer, we classify the SASI emission in our model as p-mode emission and assign a purely acoustic origin, not a vortical--acoustic origin, to it.
We compare the results of our three-dimensional model analysis with those obtained from the model's two-dimensional counterpart and find a significant reduction in the strain amplitudes in the former case, as well as significant reductions in all related quantities.
Our dominant proto-neutron star gravitational wave emission is not well characterized by emission from surface g-modes, 
complicating the relationship between peak frequencies observed and the mass and radius of the proto-neutron star expressed by  
analytic estimates under the assumption of surface g-mode emission.
We present our frequency normalized characteristic strain along with the sensitivity curves of current- and next-generation gravitational wave detectors. This simple analysis indicates that the spectrum of gravitational wave emission between $\sim$20 Hz through $\sim$1 kHz, stemming from neutrino-driven convection, the SASI, 
accretion onto the proto-neutron star, and proto-neutron star convection will be accessible for a Galactic event.
\end{abstract}



\maketitle

\section{Introduction}

The first direct detection of gravitational wave signals from a binary-black-hole merger \cite{GW150914, GW151226} opened a new era in observational astronomy and a new window on the Universe. Followed not long after by the detection of gravitational waves from a binary-neutron-star merger \cite{Abbott2017}, what was envisioned to be the importance of gravitational wave astronomy to the study of gravity {\em per se}, and astrophysics, was realized. The detections resulted in the confirmation of one of the most important predictions of general relativity -- the existence of gravitational waves. Predictions of waveforms from the inspiral of binary black holes, long sought after and finally obtained by the numerical relativity community, were validated as well. And, last but not least, confirmation that binary neutron star mergers may in fact be key to the production of heavy elements rounded out the early successes. While such confirmations reassured the relativity and astrophysics communities of their progress, the detections of the gravitational waves from these systems also raised many questions -- e.g., how are black holes of the masses inferred from the first observed black-hole inspiral produced? 

This has set the stage to prepare, even more fervently, for future detections, especially of one of the last of the primary sources of gravitational waves that will be detectable by current-generation gravitational wave detectors: a (Galactic) core-collapse supernova explosion. Core collapse supernovae are physics rich, with many processes operating in conjunction to produce a supernova. Supernova models are, therefore, innately complex. In the case of a Galactic event, a gravitational wave detection is possible \cite{Gossan16}. Such a detection, along with a detection of the supernova neutrinos,
would provide {\em direct}  information about these processes and the supernova \textquoteleft central engine\textquoteright, in turn allowing us to validate our models and to derive a better understanding of the central engine's physics. Progress in core collapse supernova theory will be necessary to address questions such as the question raised earlier regarding the masses of black holes formed during stellar core collapse. Stellar mass black holes are produced in core collapse supernovae. Theory will need to progress to better determine the outcomes of stellar collapse across the range of progenitor mass and progenitor characteristics (e.g., metallicity, rotation) observed in Nature.

\begin{figure}[ht!]
\centering
\includegraphics[width=80mm]{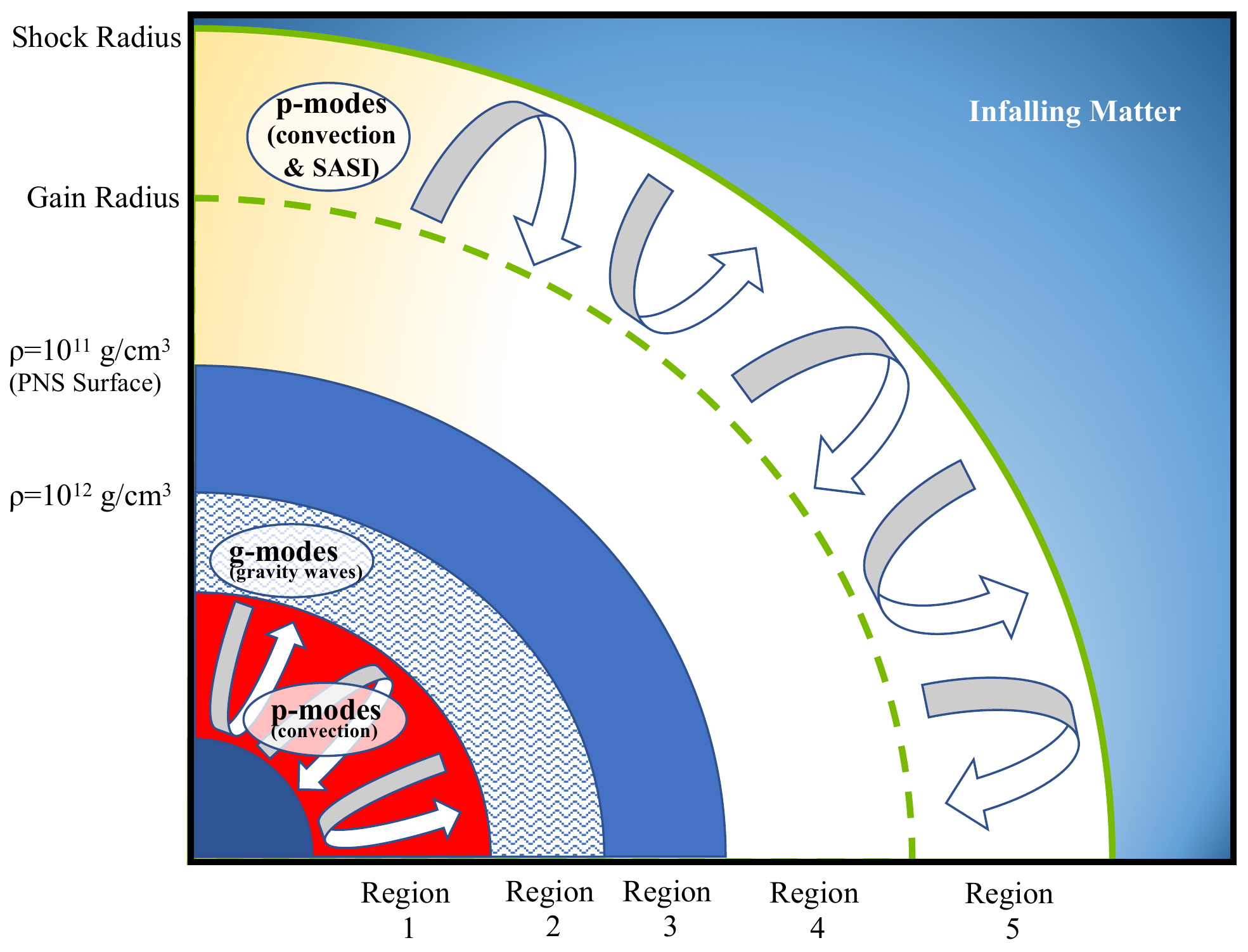}
\caption{Schematic showing the 5 critical regions in our analysis.}
\label{fig:Regions1-5Cartoon}
\end{figure}

Many studies of gravitational wave emission in core collapse supernovae based on a variety of two- and three-dimensional core collapse supernova models were performed in the past \cite{Marek09, Murphy09, Ott09, Ott12, Muller13, Kotake09, Kotake13, Scheidegger10, Fryer11, Abdikamalov14, Kuroda14, Hayama2015, Kuroda2016a, Andresen2017, Sotani2017, Kuroda2017, Morozova2018, Takiwaki2018, Hayama2018, Kawahara2018, Radice2019, Andresen2019}, including our studies \cite{Yakunin10, Yakunin15}. Arguably, progress on multidimensional core collapse supernova modeling has been exponential, in light of the increasingly powerful computational resources available to modelers, culminating in the recent three-dimensional modeling efforts of a number of groups \cite{Kuroda12, Couch13, Hanke13, Ott13, Tamborra13, Nakamura14, Takiwaki14, Tamborra14, Couch15, Foglizzo15, Lentz15, Melson15a, Melson15b, Meuller15, Kuroda2016b, Roberts16, Takiwaki2016, Mueller2017, Kuroda2018, O'Connor2018, Ott2018, Summa2018, Vartanyan2018, Burrows2019, Nakamura2019, Powell2019}. 
Studies of the gravitational wave emission based on simulation data from these latter studies have demonstrated that emission predictions made in the context of two-dimensional, axisymmetric models differ quantitatively and qualitatively from those made in the context of three-dimensional models. Moreover, conclusions regarding the physical origin of the gravitational radiation produced differ as well, especially for the dominant contributions from late-time proto-neutron star gravitational wave emission, which dominates the emission in all models, both two- and three-dimensional. Andresen et al. \cite{Andresen2017} find that gravitational wave emission in their three-dimensional models is dominated by emission from the convectively-stable, overshoot layer (Region 2 in Figure \ref{fig:Regions1-5Cartoon}, their counterpart is labelled A$_2$) above the layer of ongoing proto-neutron star convection (Region 1, their counterpart is labelled A$_1$). 
While the dominant emission of gravitational radiation still emanates from the proto-neutron star, the excitation mechanism of the modes generating the radiation is fundamentally different. In two-dimensions, the modes are excited from above. Accretion funnels resulting from neutrino-driven convection and the standing accretion shock instability (SASI) impinge on the proto-neutron star surface layers (Region 3, 
Andresen et al.'s counterpart is labelled B) and excite g-modes within them. In three dimensions, Andresen et al. find that the modes are excited internally, from below, by proto-neutron star convection. Moreover, they find a significant reduction in the amplitude of the gravitational waves emitted relative to the amplitudes they and others obtain in two dimensions. In more recent studies of gravitational wave emission based on three-dimensional models, O'Connor and Couch \cite{O'Connor2018}, Radice et al. \cite{Radice2019}, and Powell and M\"{u}ller \cite{Powell2019} come to a different conclusion. In their models $g$-mode oscillations of the convectively stable layers below the surface of the proto-neutron star remained excited from above, as in the two-dimensional case. Moreover, for Radice et al., the dominant gravitational wave emission after $\sim$400 ms of post-bounce evolution stemmed from the fundamental (quadrupolar) mode, not from $g$-modes. In our study, we expand the spectrum of possibilities.

Thus, gravitational wave analysis in the context of some of today's most sophisticated three-dimensional core collapse supernova models is so far pointing to a model-dependency to some of the critical aspects -- specifically, the source and nature -- of the gravitational wave emissions. As additional three-dimensional models become available -- particularly in the context of equivalent models across different groups using different codes -- we will be able to separate out any biases introduced by numerical effects. But fundamental {\em physical} differences for different equations of state and/or different progenitors and progenitor characteristics may remain. Along with continued three-dimensional modeling and the resultant growth in the number of models available, the emergent field of proto-neutron star seismology \cite{Sotani2019,Torres-Forne19,Westernacher-Schneider2019} will be useful to help sort through the possibilities, to properly classify the modes leading to gravitational wave emission in the community's emerging three-dimensional models. This will be essential to extracting core collapse supernova and proto-neutron star physics when we are fortunate enough to detect the gravitational waves from a (Galactic) event.

\section{Models and Methods}
Our gravitational wave analysis is based on the data generated in the two- and three-dimensional core collapse supernova simulations performed by \citet{Lentz15}. 
The simulations were both initiated from the non-rotating 15\msun progenitor of \citet{Woosley07}. 
The simulations were performed with the \chimera code, which is based on multigroup flux-limited diffusion in the ray-by-ray approximation, Newtonian self-gravity with a monopole correction to account for the effects of general relativity, Newtonian hydrodynamics, and a nuclear reaction network \cite{bruenn2018chimera}. \chimera includes electron capture on protons and nuclei, the latter using the LMSH capture rates, electron--positron annihilation, and nucleon--nucleon bremsstrahlung, along with their inverse weak interactions. It also includes coherent isoenergetic scattering on nuclei, as well as neutrino--electron (large-energy transfer) and neutrino--nucleon (small-energy transfer) scattering. 
In both simulations, we employed two equations of states: \citet{Lattimer91} (with an incompressibility $K = 220$~MeV) for $\rho > 10^{11}$~\gcc and an enhanced version of the \citet{Cooperstein85} equation of state for $\rho < 10^{11}$~\gcc. In outer regions, we employed a 14-species $\alpha$-network \cite{Hix99}.

The three-dimensional computational grid for model C15-3D comprised 540($r$)$\times$180($\theta$)$\times$180($\phi$) zones equally distributed in the $\phi$-direction only. (Here, ``C'' means from the \chimera C-Series simulation suite, and 15 delineates the progenitor mass.) 
The $\theta$-resolution in the three-dimensional model varied from $ 2/3^\circ$ near the equator to $8.5^\circ$ near the poles (i.e., to keep $\mu\equiv\cos\theta$ constant). 
The $\theta$-resolution in the two-dimensional model, C15-2D, was uniformly $0.7^\circ$. 
The radial resolution in both simulations varied according to conditions of the moving grid and  reached  0.1~km inside the PNS. 
Model C15-3D was evolved in 1D during collapse and through bounce. At 1.3 ms after bounce random density perturbations of 0.1\% were applied to the matter between 10--30 km, which is the region that had been shocked.

We employ the quadrupole approximation for extracting the gravitational wave signals from the mass motions, using the expressions detailed in \cite{Yakunin15} (for the two-dimensional case) and below (for the three-dimensional case). Unless otherwise noted, the results presented here are for model C15-3D.
To isolate the impact of dimensionality, all comparisons of the gravitational wave emissions in C15-2D and C15-3D have been performed using the same evolution time frame, dictated by the duration of both runs (0--450~ms).

We begin with the lowest multipole (quadrupole) moment of the Transverse-Traceless gravitational wave strain \cite{Kotake2006}
\begin{equation}
h_{ij}^{\rm TT}=\frac{G}{c^4}\frac{1}{r}\sum_{m=-2}^{+2}\frac{d^{2}I_{2m}}{dt^{2}}(t-\frac{r}{c})f_{ij}^{2m},
\label{eq:quadrupoleexpansion}
\end{equation}

\noindent where $i$ and $j$ run over $r$, $\theta$, and $\phi$ and where $f_{ij}^{2m}$ are the tensor spherical harmonics, given by

\begin{equation}
f_{ij}^{2m} = \alpha r^{2}
\begin{pmatrix}
0 & 0            & 0                                  \\
0 & W_{2m} & X_{2m}                         \\
0 & X_{2m}  & -W_{2m}\sin^{2}\theta  \\
\end{pmatrix},
\label{eq:tensorsphericalharmonics}
\end{equation}

\noindent with

\begin{equation}
X_{2m}=2\frac{\partial}{\partial\phi}\left(\frac{\partial}{\partial\theta}-\cot\theta\right) Y_{2m}(\theta,\phi)
\label{eq:Xlm}
\end{equation}

\noindent and 

\begin{equation}
W_{2m}=\left(\frac{\partial ^2}{\partial \theta^2}-\cot\theta\frac{\partial}{\partial\theta}-\frac{1}{\sin^2\theta}\frac{\partial^2}{\partial\phi^2}\right)Y_{2m}(\theta, \phi).
\label{eq:Wlm}
\end{equation}

\noindent The normalization, $\alpha$, is determined by

\begin{equation}
 \int d\Omega \left(f_{l m}\right)_{ab}
\left(f_{l' m'}^*\right)_{cd}
\gamma^{ac}\gamma^{bd}
= r^4 \delta_{l l'}\delta_{m m'},
\end{equation}

\noindent where $a,b,c,d=\theta,\phi$, and $\gamma_{ab}$ is the 2-sphere metric

\begin{equation}
 \gamma_{ab} = 
  \begin{bmatrix}
   1 & 0 \\
   0 & \sin^2\theta
  \end{bmatrix}.
\end{equation}

\noindent For $l=2$,  $\alpha = \frac{1}{4\sqrt{3}}$.

The mass quadrupole is

\begin{equation}
I_{2m}=\frac{16\sqrt{3}\pi}{15} \int \tau_{00}Y_{2m}^{*}r^2dV.
\label{eq:massquadrupole}
\end{equation}

\noindent In equation (\ref{eq:massquadrupole}), $dV=r^{2}\sin\theta dr d\theta d\phi$, and $\tau_{00}$ is simply the rest-mass density, $\rho$, for the weak fields assumed here.
(N.B. The coefficient $\frac{1}{15}$ was incorrectly written as $\frac{1}{5}$ in \cite{Yakunin10,Yakunin15}.) We also define the gravitational wave amplitude

\begin{equation}
A_{2m}\equiv\frac{G}{c^4}\frac{d^{2}I_{2m}}{dt^{2}}.
\label{eq:gravwaveamplitude}
\end{equation}

\noindent Any gravitational wave extraction method should endeavor to minimize numerical noise. Unfortunately, most numerical differentiation methods amplify numerical noise. To avoid this, we define

\begin{equation}
A_{2m}\equiv\frac{dN_{2m}}{dt},
\label{eq:N2mdot}
\end{equation}

\noindent where

\begin{equation}
N_{2m}=\frac{G}{c^4}\frac{dI_{2m}}{dt}.
\label{eq:N2mdot}
\end{equation}

\noindent Combining equations (\ref{eq:massquadrupole}) and (\ref{eq:N2mdot}), we obtain

\begin{eqnarray}
\label{eq:N2mdot2}
N_{2m} & = & \frac{16\sqrt{3}\pi G}{15c^4}\frac{d}{dt} \int \rho Y_{2m}^{*}r^2dV \\ \nn
& = & \frac{16\sqrt{3}\pi G}{15c^4}\int \frac{\partial \rho}{\partial t}Y_{2m}^{*}r^2dV. \nn
\end{eqnarray}

\noindent The continuity equation can be used to eliminate the time derivative in equation (\ref{eq:N2mdot2}), which gives \cite{Finn90}

\begin{eqnarray}
\label{eq:n2m-integration}
 N_{2m}& = &\frac{16\sqrt{3}\pi G}{15c^4}
 \int_0^{2\pi}d\varphi' \int_0^\pi d\vartheta' \int_0^\infty dr'~r'^3 \\ \nn
  && \left[
2\rho v^{\hat{r}}Y^*_{2m}\sin\vartheta'
+ \rho v^{\hat{\vartheta}}\sin\vartheta'
\frac{\partial}{\partial\vartheta'}Y^*_{2m} \right. \\ \nn
&& \left.+ \rho v^{\hat{\varphi}}
\frac{\partial}{\partial\varphi'}Y^*_{2m}
\right], \nn
\end{eqnarray}
where $r'$, $\vartheta'$ and $\varphi'$ are the spherical coordinates in the source frame. $v^{i'}$ are the components of the velocity in the same frame. (N.B. The factor $\frac{1}{15}$ is missing in \cite{Yakunin15}.)
Finally, to compute the gravitational wave amplitude, equation (\ref{eq:gravwaveamplitude}), we evaluate the time derivative of $N_{2m}$ numerically by computing $N_{2m}$ using equation (\ref{eq:n2m-integration}) on each time slice of our simulation and in turn computing the time derivative by differencing the values of $N_{2m}$ obtained on adjacent slices using a second-order finite-difference stencil. Finally, we compute the gravitational wave strains for both polarizations, which are related to $h^{TT}_{ij}$ by:

\begin{eqnarray}
\label{eq:rhpus}
h_+ & = & \frac{h^{TT}_{\theta\theta}}{r^2}, \\ 
h_{\times} & = & \frac{h^{TT}_{\theta\phi}}{r^2 \sin\theta}. \\ \nonumber
\end{eqnarray}

The total luminosity emitted in gravitational waves is given by \cite{Thorne80}

\begin{equation}
\frac{dE}{dt}=\frac{c^3}{G}\frac{1}{32\pi}\sum_{m=-2}^{+2}\langle\left|\frac{dA_{2m}}{dt}\right|^2\rangle,
\label{eq:gravwaveluminosity}
\end{equation}

\noindent where the $\langle\rangle$ indicate averaging over several wave cycles. To compute the spectral signatures, we must relate the gravitational wave  luminosity to its spectrum, using Parseval's Theorem:

\begin{equation}
\int_{-\infty}^{+\infty}|x(t)|^2dt=\int_{-\infty}^{+\infty}|\tilde{x}(2\pi f)|^2df.
\label{eq:ParsevalsTheorem}
\end{equation}

\noindent Here, $\tilde{x}(2\pi f)$ is the Fourier transform of $x(t)$. The total energy emitted in gravitational waves is 

\begin{eqnarray}
\label{eq:totalGWenergyf}
E=\int_{-\infty}^{+\infty}\frac{dE}{dt}dt & = & \frac{c^3}{32\pi G}\sum_{m=-2}^{+2}\int_{-\infty}^{+\infty}|\dot{A}_{2m}|^2dt \\ \nonumber
& = &\frac{c^3}{32\pi G}\sum_{m=-2}^{+2}\int_{-\infty}^{+\infty}|\tilde{\dot{A}}_{2m}(2\pi f)|^2df \\ \nonumber
& = &\frac{c^3}{16\pi G}\sum_{m=-2}^{+2}\int_{0}^{+\infty}|\tilde{\dot{A}}_{2m}(2\pi f)|^2df. \\ \nonumber
\end{eqnarray}

\noindent where the over-dot now represents the time derivative. The time derivative of $\tilde{A}_{2m}$ in equation (\ref{eq:totalGWenergyf}) can be eliminated using the standard property of Fourier transforms -- i.e.,

\begin{equation}
|\tilde{\dot{A}}_{2m}(2\pi f)|^2=(2\pi f)^2|\tilde{A}_{2m}(2\pi f)|^2.
\label{eq:FTproperty}
\end{equation}

\noindent Inserting equation (\ref{eq:FTproperty}) in equation (\ref{eq:totalGWenergyf}) and taking the derivative with respect to frequency gives

\begin{equation}
\frac{dE}{df}=\frac{c^3}{16 \pi G}(2\pi f)^2\sum_{m=-2}^{+2}|\tilde{A}_{2m}|^2.
\label{eq:dedf}
\end{equation}

\noindent The stochastic nature of GW signals from core collapse supernovae prompts the use of short-time Fourier transform (STFT)
techniques to determine $\tilde{A}_{2m}$ \cite{Murphy09}:
\begin{equation}
  \textrm{STFT}\{A_{2m}(t)\}\left(\tau, f\right) =  \int\limits_{-\infty}^{\infty}
            A_{2m}(t)\,H(t - \tau)e^{-i\,2\pi ft}dt
\end{equation}
where $H(t - \tau)$ is the Hann window function.
In our analysis, we set the window width to $\sim$15 ms. The sampling interval of our data is $\sim$0.02 ms 
during the first tens of ms after bounce, rises to $\sim$0.15 ms at 100 ms, and then settles down to a value 
$\sim$0.10 ms from 300 ms post bounce until the end of our run. Data from this non-uniform temporal grid is interpolated
onto a uniform temporal grid prior to the computation of the short-time Fourier transform.
Finally, we relate $dE/df$ to the characteristic gravitational 
wave strain, defined by \cite{Flanagan98}

\begin{equation}
h^{2}_{\rm char}(f)=\frac{2G(1+z)^2}{\pi^2 c^3 D^2(z)}\frac{dE}{df}[(1+z)f],
\label{eq:hchar}
\end{equation}

\noindent where $z$ is the source's redshift. Here we assume $z=0$, as for a Galactic supernova. Then equation (\ref{eq:hchar}) becomes

\begin{equation}
h_{\rm char}(f)=\sqrt{\frac{2G}{\pi^2 c^3 D^2}\frac{dE}{df}}.
\label{eq:hchar3}
\end{equation}

\section{Results}

\begin{figure}[ht!]
\centering
\includegraphics[width=80mm]{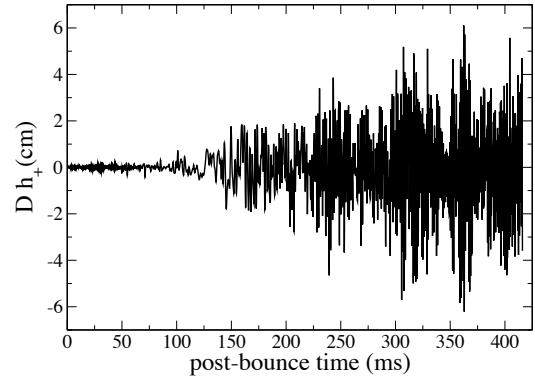}
\caption{
The gravitational wave strain for the $+$ polarization, plotted as a function of time over the course of our simulation, 
viewed along the $z$-axis. The period between bounce and $\sim$100 ms is fairly quiescent with regard to gravitational wave emission. This 
initial quiescent period is followed by a period during which neutrino-driven convection and the SASI develop, generating low-frequency 
($<$200 Hz) gravitational waves for the duration of the simulation. In the period between $\sim$150--200 ms, this low-frequency emission is joined by intermediate 
frequency emission from the proto-neutron star, in the range $\sim$400--600 Hz, due to neutrino-driven convection- and SASI-induced 
aspherical accretion onto it. After $\sim$200 ms, gravitational wave emission is dominated by high-frequency emission, above $\sim$600 Hz, 
by a second phase of Ledoux convection deep within the proto-neutron star, which is long-lived and persists to the end of our simulation, as well. 
The change in character of the gravitational wave strain across these phases is readily seen in the plot.
}
\label{fig:rh+fig}
\end{figure}

\begin{figure}[ht!]
\centering
\includegraphics[width=80mm]{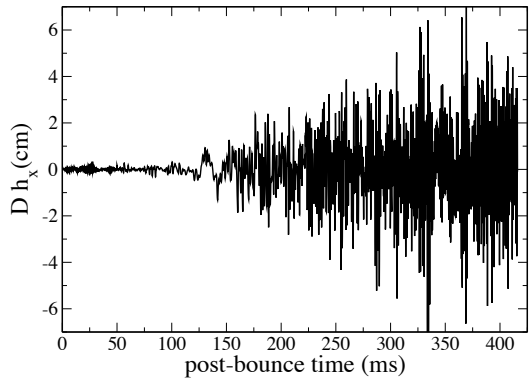}
\caption{Same as in Figure \ref{fig:rh+fig} but for the $\times$ polarization.}
\label{fig:rhxfig}
\end{figure}

\begin{figure}[ht!]
\centering
\includegraphics[width=80mm]{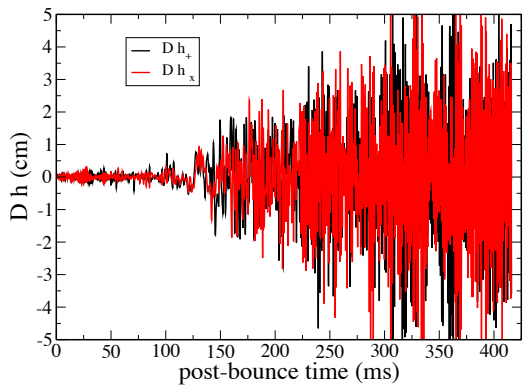}
\caption{
Plot of the strain amplitude for both polarizations, viewed along the $z$-axis. Clearly the strain amplitudes for the two polarizations are comparable, which reflects the fact there is no preferred physical direction in our model.
}
\label{fig:rh+rhx_zaxis}
\end{figure}

\begin{figure}[ht!]
\centering
\includegraphics[width=80mm]{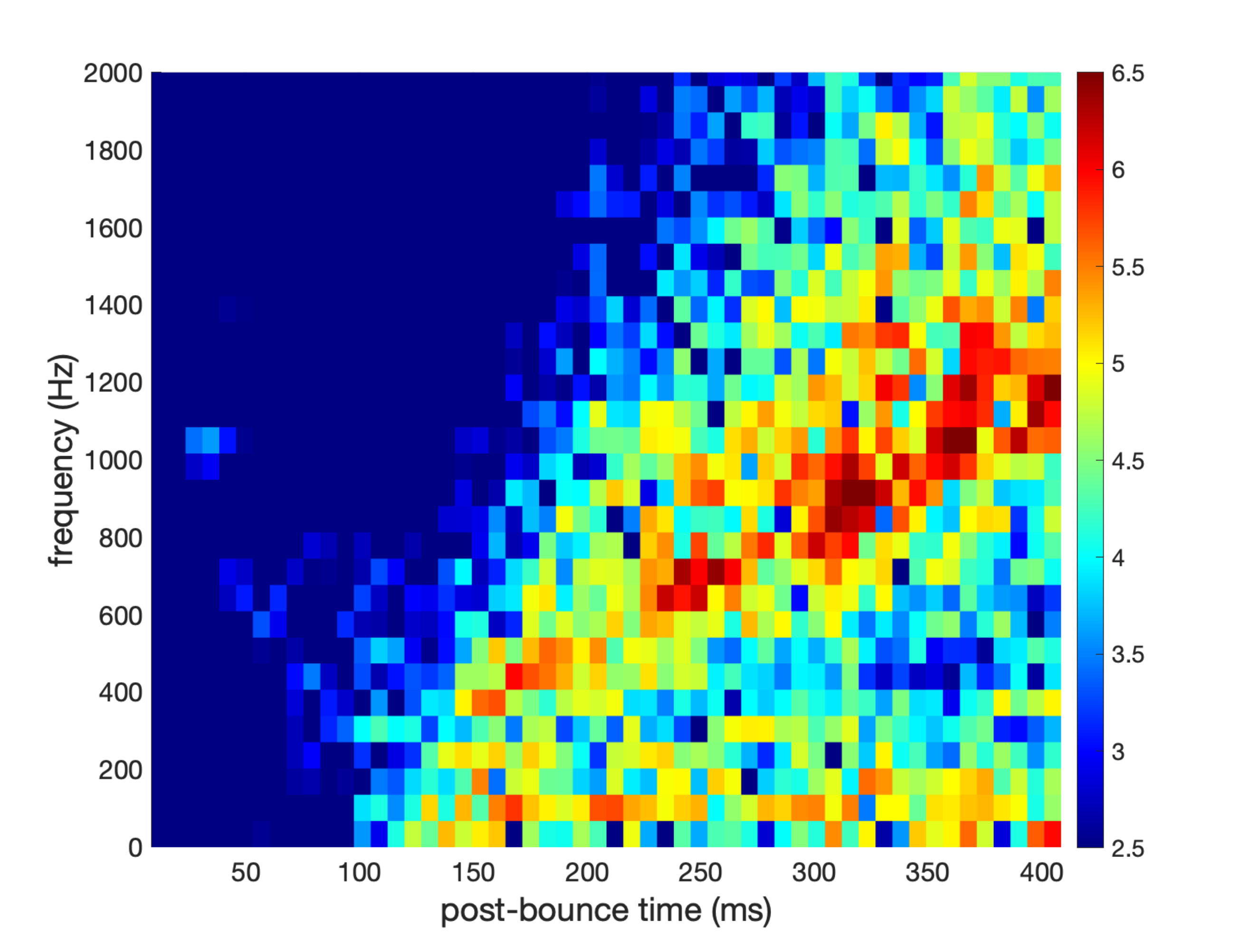}
\caption{
The Fourier transform of $h_+$ binned in frequency and time over the course of our simulation. Three distinct features are evident. After $\sim$100 ms, gravitational radiation with frequencies less than $\sim$200 Hz is emitted, due to aspherical mass motions in the gain layer from neutrino-driven convection and the SASI. Between 
$\sim$150--200 ms, this is joined by intermediate-frequency emission from the proto-neutron star, in the range $\sim$400--600 Hz, due to aspherical accretion onto it. The accretion flows become aspherical after neutrino-driven convection and the SASI develop. After $\sim$200 ms, gravitational radiation at high frequencies $>$600 Hz is emitted, due to Ledoux convection in the proto-neutron star. The peak frequency of this emission rises as the proto-neutron star evolves.
}
\label{fig:heatmapfig}
\end{figure}

\subsection{Temporal Analysis}

Figures \ref{fig:rh+fig}, \ref{fig:rhxfig}, and \ref{fig:heatmapfig} show the time evolution of $Dh_+$, $Dh_\times$, and the Fourier transform of 
$h_+$, respectively, as a function of time after bounce, for the entire duration of our run. (Here, $D$ is the distance to the supernova.)
After the quiescent phase, which lasts until $t\sim100$ ms after bounce, a signal with frequencies below $\sim$200 Hz begins, which persists through the remainder of our simulation. This phase of gravitational wave emission corresponds to the development of aspherical mass motion in the gain layer, or net neutrino heating layer, between the gain radius and the shock, due to the development of neutrino-driven convection there and the development of the standing accretion shock instability (SASI). This low-frequency emission persists throughout the remainder of our run, as neutrino-driven convection and the SASI are maintained through the entire remaining evolution we cover in this model.
After $\sim$150 ms after bounce, the low-frequency signal is joined by another, intermediate-frequency signal, between $\sim$400--600 Hz. This is particularly evident in Figure \ref{fig:heatmapfig} between 150 and 200 ms after bounce. We associate this with gravitational wave emission by the proto-neutron star (mostly from Regions 2 and 3; see Figures \ref{fig:heatmap2} and \ref{fig:heatmap3} below) as the aspherical accretion flow impinges on it once neutrino-driven convection and the SASI have developed. This component of the gravitational wave emission weakens as the shock radius begins to expand and explosion is initiated, which occurs after $\sim$200 ms in this model \cite{Lentz15}. After $\sim$200 ms, the final and dominant phase of gravitational wave emission begins. This is due to a second, but now long-lived phase of Ledoux convection deep in the interior of the proto-neutron star. Unlike the case of early prompt Ledoux convection in the proto-neutron star, this second phase of Ledoux convection is sustained by continued neutrino diffusion out of the core and, consequently, by the maintenance of the lepton gradients that drive it. The rise in the peak frequency for this higher-frequency branch of the gravitational wave emission results from the evolution of the proto-neutron star as it deleptonizes and contracts. 
In Figure \ref{fig:rh+rhx_zaxis}, we plot the strains associated with the two polarizations. They are clearly comparable in magnitude and share very similar time dependence, which mirrors the hydrodynamics we observe in this model. No particular direction can be singled out (although, along these lines, we comment on the impact of our constant-mu grid in Section \ref{sec:numericalconsiderations}).

\begin{figure}[ht!]
\centering
\includegraphics[width=80mm]{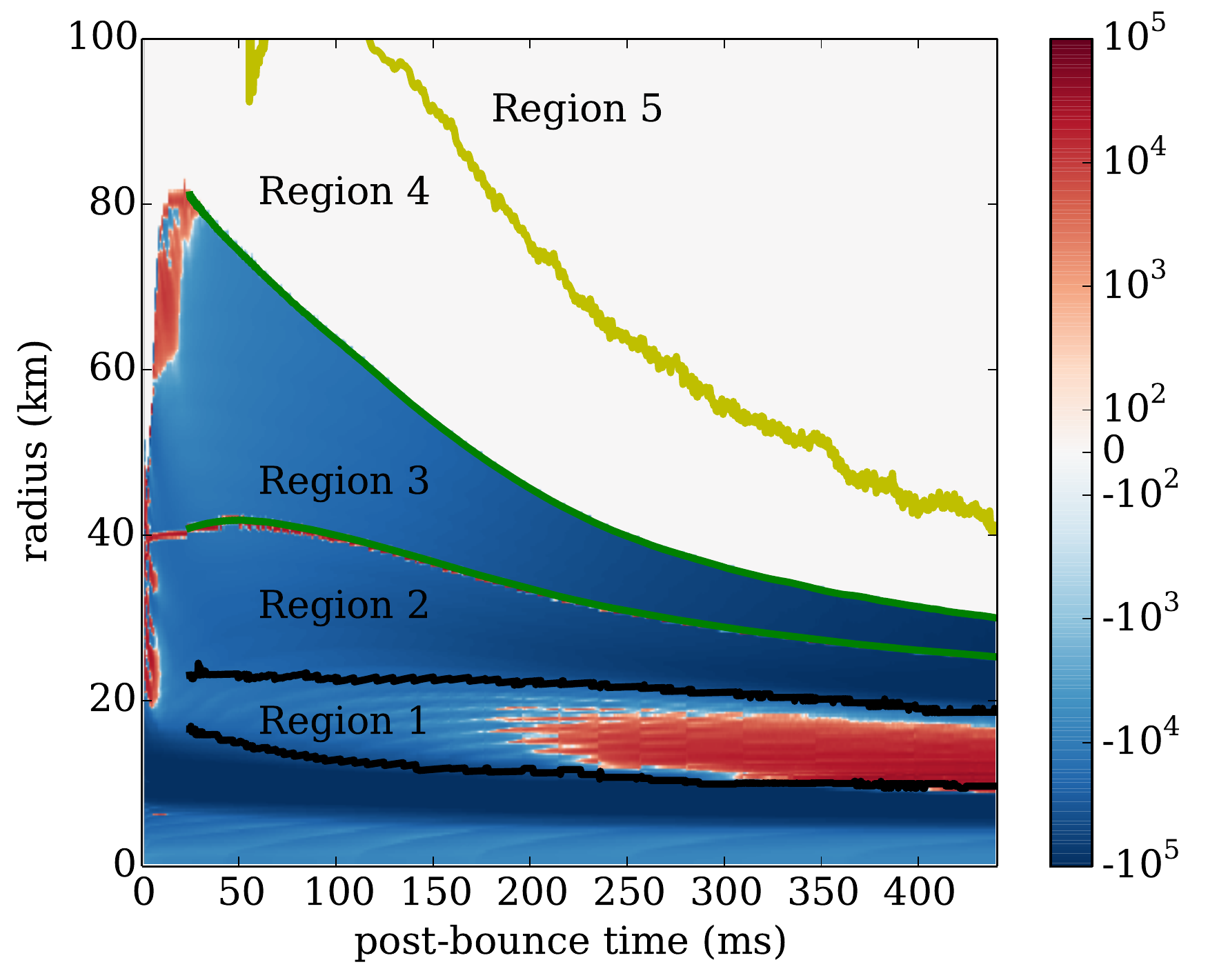}
\caption{Plot of the shell-averaged Brunt--V\"{a}is\"{a}l\"{a} (BV) frequency as a function of radius and time in our simulation. The scale on the right-hand side is in Hz. Stability (instability) to Ledoux convection is indicated by blue (red) shading. Five contours are pronounced in the plot, bounding five regions. Two contours (black, located where the convective velocities are 5\% of peak) bound the region of convective overturn deep within the proto-neutron star. Moving outward, two contours (dark green) mark the $\rho = 10^{12,11}$ g\, cm$^{-3}$, constant-density contours, respectively. The outermost contour (light green) traces the angle-averaged gain radius. In our model, the surface of the proto-neutron star is defined by $\rho = 10^{11}$ g\, cm$^{-3}$, above which the BV frequency is not computed. Early Ledoux instability between 20 and 50 km and for postbounce times less than $\sim$10 ms is indicated, as well as instability between 60 and 80 km up to $\sim$30 ms after bounce. Deep proto-neutron star Ledoux instability is evident, as well, beginning after $\sim$175 ms after bounce and continuing for the duration of our simulation, between $\sim$9 km and 20 km.}
\label{fig:BVfig}
\end{figure}

\begin{figure}[ht!]
\centering
\includegraphics[width=80mm]{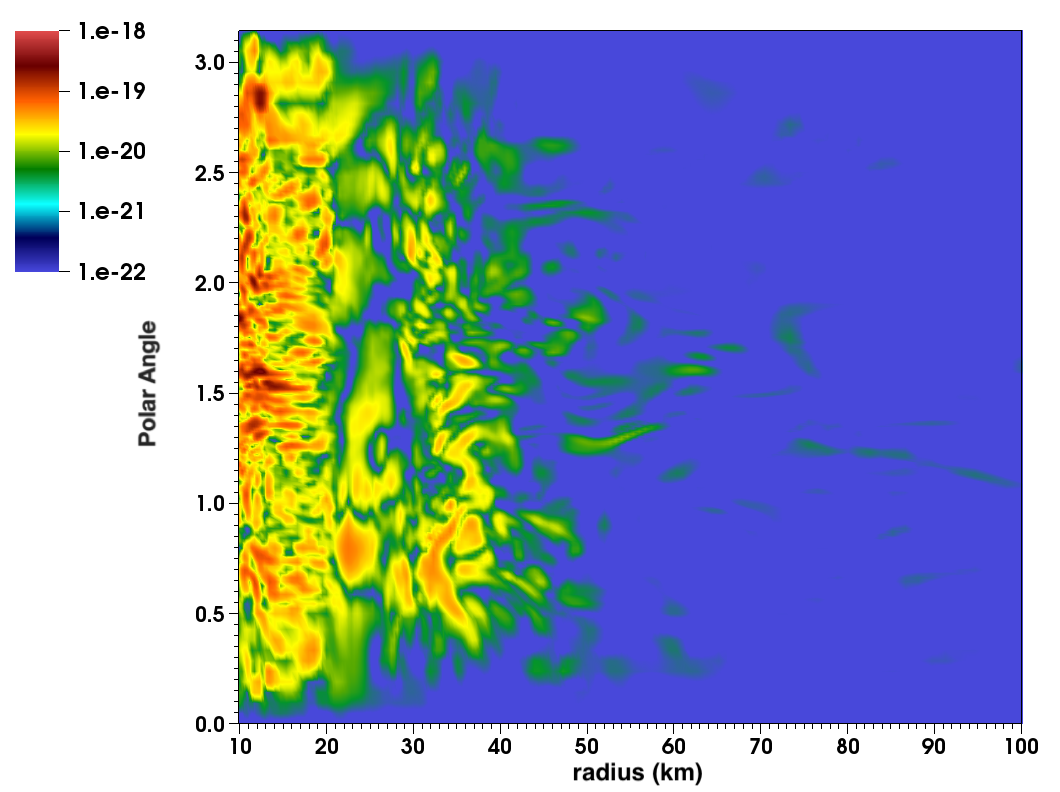}
\caption{The integrand (Equation (\ref{eq:n2m-integration20})) of the gravitational wave amplitude $A_{20}$ is plotted for $\phi=0$ as a function of $r$ and $\theta$ at a time $\sim400$ ms after bounce. The largest amplitudes are seen concentrated in the region between 10 and 20 km in radius, for all $\theta$. We attribute these amplitudes to Ledoux convection in this region of the proto-neutron star, which begins after $\sim$175 ms after bounce and persists throughout our simulation. Nontrivial amplitudes are also evident in the region just below and above the proto-neutron star surface, which is located at $\sim$30 km at this time. These latter amplitudes are induced by a combination of convective overshoot (undershoot) from the regions below (above).}
\label{fig:dN20dtphi0}
\end{figure}

\begin{figure}[ht!]
\centering
\includegraphics[width=80mm]{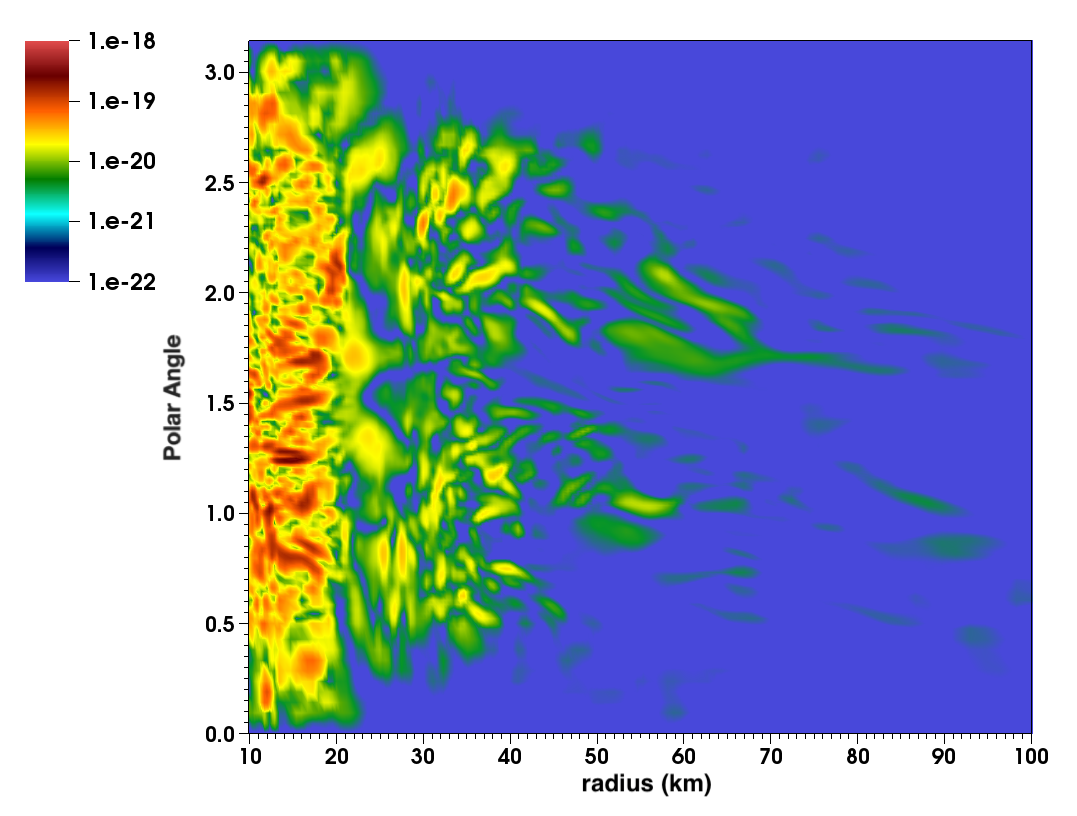}
\caption{Same as in Figure \ref{fig:dN20dtphi0} but for $\phi = \pi/2$.
In this snapshot, gravitational wave amplitudes are particularly evident in the region above $\sim$45 km, signatures of the aspherical mass motions in the gain layer.}
\label{fig:dN20dtphipi2}
\end{figure}

\begin{figure}[ht!]
\centering
\includegraphics[width=80mm]{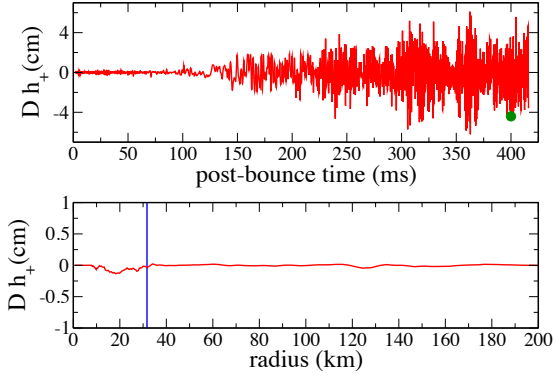}
\caption{Top Panel: The total gravitational wave strain for the $+$ polarization at $\sim$399.97 ms after bounce is marked by a green dot. Bottom Panel: The $+$ polarization gravitational wave strain is given as a function of radius only, now integrated over $\theta$ relative to what was shown in Figures \ref{fig:dN20dtphi0} and \ref{fig:dN20dtphipi2}
for $A_{20}$, but now including all contributions, $A_{2m}$, for $m\neq0$. The vertical blue line marks the proto-neutron star surface. The total gravitational wave strain at this time is clearly dominated by contributions from the region between 10 and 20 km, where there is ongoing Ledoux convection.}
\label{fig:2panelfig399}
\end{figure}

\begin{figure}[ht!]
\centering
\includegraphics[width=80mm]{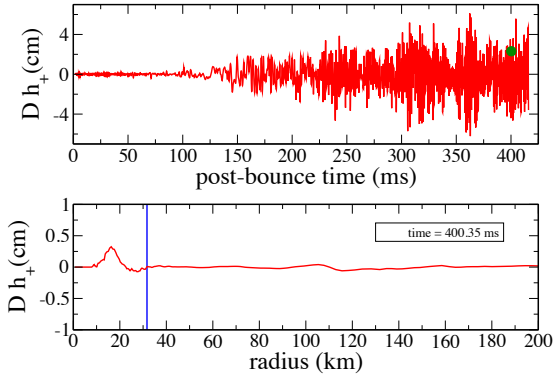}
\caption{Same as in Figure \ref{fig:2panelfig399} but 380 $\mu$s later. Here we demonstrate the rapid variability of the strain in the Ledoux convecting region.}
\label{fig:2panelfig400}
\end{figure}

Figure \ref{fig:BVfig} is a plot of the shell-averaged Brunt-V\"{a}is\"{a}l\"{a} (BV) frequency as a function of radius and post-bounce time over the course of our run. The BV frequency plotted here is given by \cite{Bruenn2004}

\begin{equation}
\omega_{\rm BV}=\sqrt{
-\frac{1}{\rho}\frac{\partial \Phi}{\partial r}
\left[
\left(\frac{\partial \rho}{\partial s}\right)_{P,Y_l}\frac{ds}{dr}+
\left(\frac{\partial \rho}{\partial Y_l}\right)_{P,s}\frac{dY_l}{dr}
\right]
}.
\label{eq:BVfreq}
\end{equation}
where $\Phi, s, Y_{l}$ are the gravitational potential, entropy per baryon, and total lepton fraction (electrons plus neutrinos), respectively.
The BV frequency is evaluated
using the \chimera\ simulation data. For $\rho\ge 10^{12}$ gcm$^{-3}$, the neutrino fraction in $Y_l$ is computed assuming the neutrino distribution function is the equilibrium distribution function at the local thermodynamic conditions. For lower densities, we substitute $Y_e$ for $Y_l$ in Equation (\ref{eq:BVfreq}).
Ledoux stable (unstable) regions correspond to real (imaginary) $\omega_{\rm BV}$.
Ledoux unstable regions are shown in red. Stable regions are shown in blue. Five contours are clearly visible. The innermost contours (shown in black, located where the convective velocities are 5\% of peak) bound the region of convective overturn deep in the proto-neutron star interior. Moving outward in radius, two additional contours (shown in dark green) mark the $10^{12}$ g\, cm$^{-3}$ and $10^{11}$ g\, cm$^{-3}$ constant-density contours, respectively. The $10^{11}$ g\, cm$^{-3}$ constant-density contour corresponds to the (defined) surface of the proto-neutron star. Finally, the outermost contour (shown in light green) marks the gain radius, bounding the gain layer from below. We do not compute the BV frequency above the proto-neutron star surface. There, the BV frequency is set to zero. The early Ledoux instabilities between 20 and 50 km and between 60 and 80 km initiate prompt convection, but by $\sim$30 ms after bounce, prompt convection renders this region of the core Ledoux stable. No significant gravitational wave emission occurs in association with this first phase of proto-neutron star Ledoux instability. However, at $\sim$175 ms after bounce, between 10 and 20 km, clearly the proto-neutron star core once again becomes Ledoux unstable and remains unstable for the duration of the run. In turn, this long-lived Ledoux instability drives convection in the region, which is responsible for the higher-frequency gravitational wave emission that persists, as well.
(The red line along the $10^{12}$ g\, cm$^{-3}$ constant-density contour is an artifact and simply the result of switching there between evaluating the BV frequency using $Y_\ell$ versus $Y_e$ in Equation (\ref{eq:BVfreq}) as we move inward toward higher densities. It does not reflect a physical Ledoux instability.)

Direct evidence that the gravitational wave emission in our model after 200 ms post bounce is dominated by Ledoux convection in the proto-neutron star is provided by looking at snapshots of the gravitational wave amplitudes at late times in the run. Figures \ref{fig:dN20dtphi0} and \ref{fig:dN20dtphipi2} show the integrand of the gravitational wave amplitude $A_{20}$ as a function of $r$ and $\theta$ for two values of $\phi$, at $\sim$400 ms after bounce. Here we plot the time derivative (computed by differencing) of the integrand in Equation (\ref{eq:n2m-integration}), for $m=0$. Specifically, we plot 

\begin{equation}
\label{eq:n2m-integration20}
8 \sqrt{\frac{\pi}{15}} \frac{G}{c^4}
\frac{\Delta
\{
r^{3}\rho\sin\theta
[
v^{r}(3\cos^{2}\theta-1)
-3v^{\theta}\sin\theta\cos\theta
]
\}
}
{\Delta t}.
\end{equation}

\noindent It is clear that the amplitude is largest in the region between 10 and 20 km, where the proto-neutron star is Ledoux unstable. The proto-neutron star radius at this time is approximately 30 km (this can be read off of Figure \ref{fig:BVfig}). 
Modest gravitational wave amplitudes in regions just below and just above the proto-neutron star surface, as well as above $\sim$45 km -- i.e., in the gain layer -- can be seen, as well.

In Figures \ref{fig:2panelfig399} and \ref{fig:2panelfig400} we plot the gravitational wave strain as a function of radius only (lower panels in both figures). The upper panels show the entire strain over the course of our run. The total strain at the particular post-bounce time when we look at the strain's radial profile is indicated by the green dot in the upper panel. The value of the strain indicated by the dot is simply the sum of all of the strains across the plot in the lower panel -- i.e., each amplitude across our radial grid in the lower panel is the value of the gravitational wave strain for its radial shell, including both the integrand and the volume element in Equation (\ref{eq:n2m-integration}).
The vertical blue line in both plots marks the radius of the proto-neutron star surface. Two time slices are displayed. In both cases, it is evident that the strain is largest between 10 and 20 km -- i.e., in the Ledoux convective region. Also evident by comparing the strain in this region in both figures is its rapid variability, with the magnitude and even sign of the radial profile of the strain changing rapidly over a period of only $\sim$380 $\mu$s.

\begin{figure}[ht!]
\centering
\includegraphics[width=80mm]{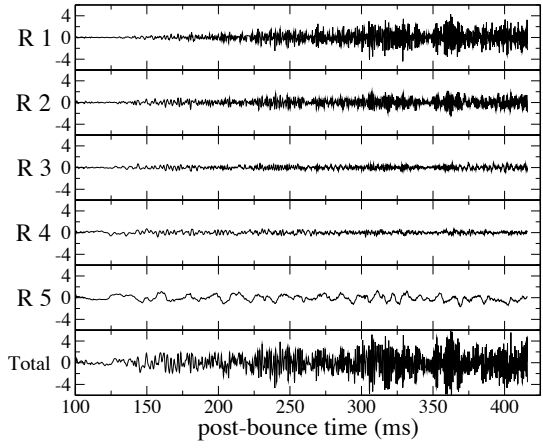}
\caption{Same as in Figure \ref{fig:rh+fig} but decomposed in terms of the 5 layers defined in Figure \ref{fig:BVfig}.}
\label{fig:rh+bylayers}
\end{figure}

\begin{figure}[ht!]
\centering
\includegraphics[width=80mm]{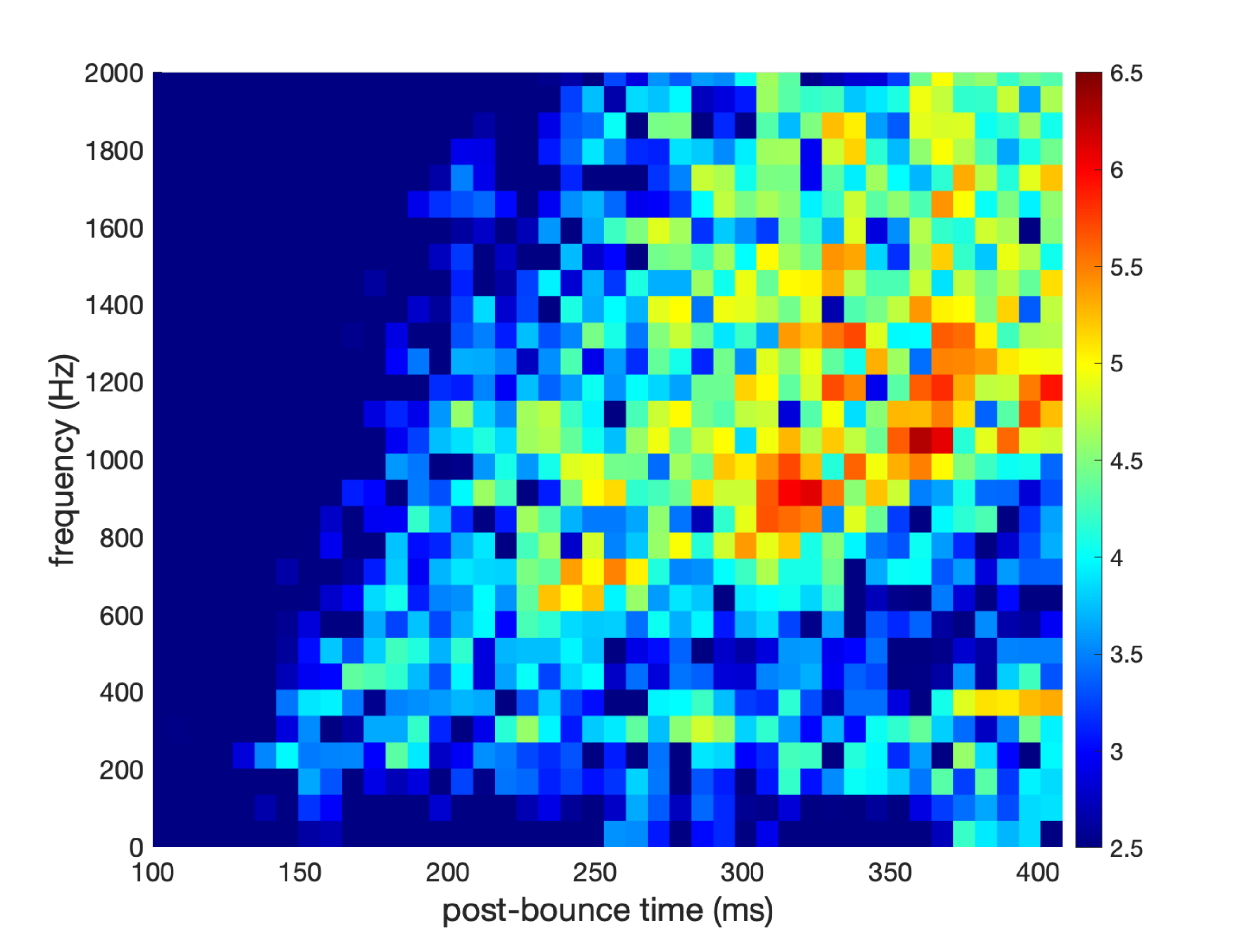}
\caption{Same as in Figure \ref{fig:heatmapfig} but for Region 1.}
\label{fig:heatmap1}
\end{figure}

\begin{figure}[ht!]
\centering
\includegraphics[width=80mm]{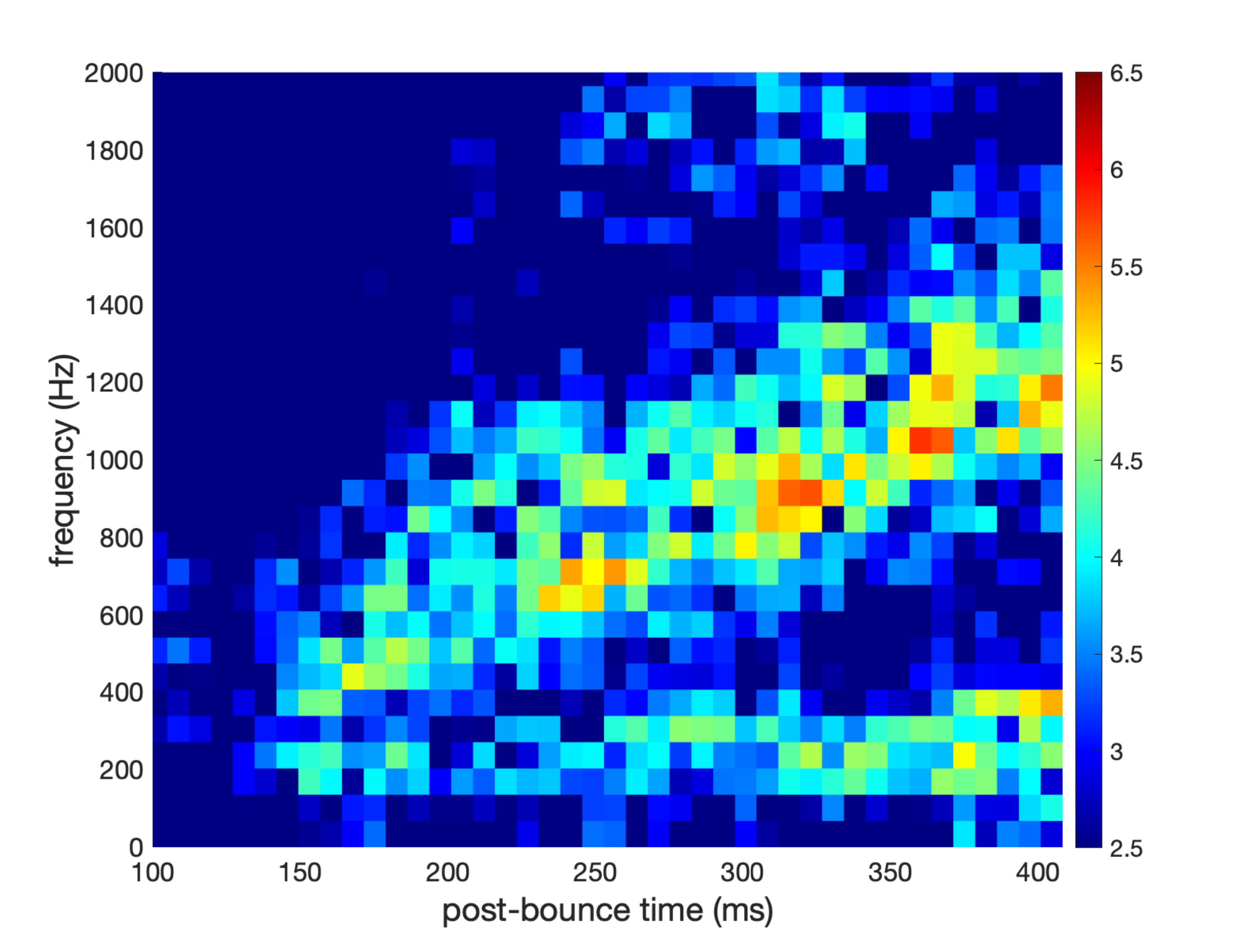}
\caption{Same as in Figure \ref{fig:heatmapfig} but for Region 2.}
\label{fig:heatmap2}
\end{figure}

\begin{figure}[ht!]
\centering
\includegraphics[width=80mm]{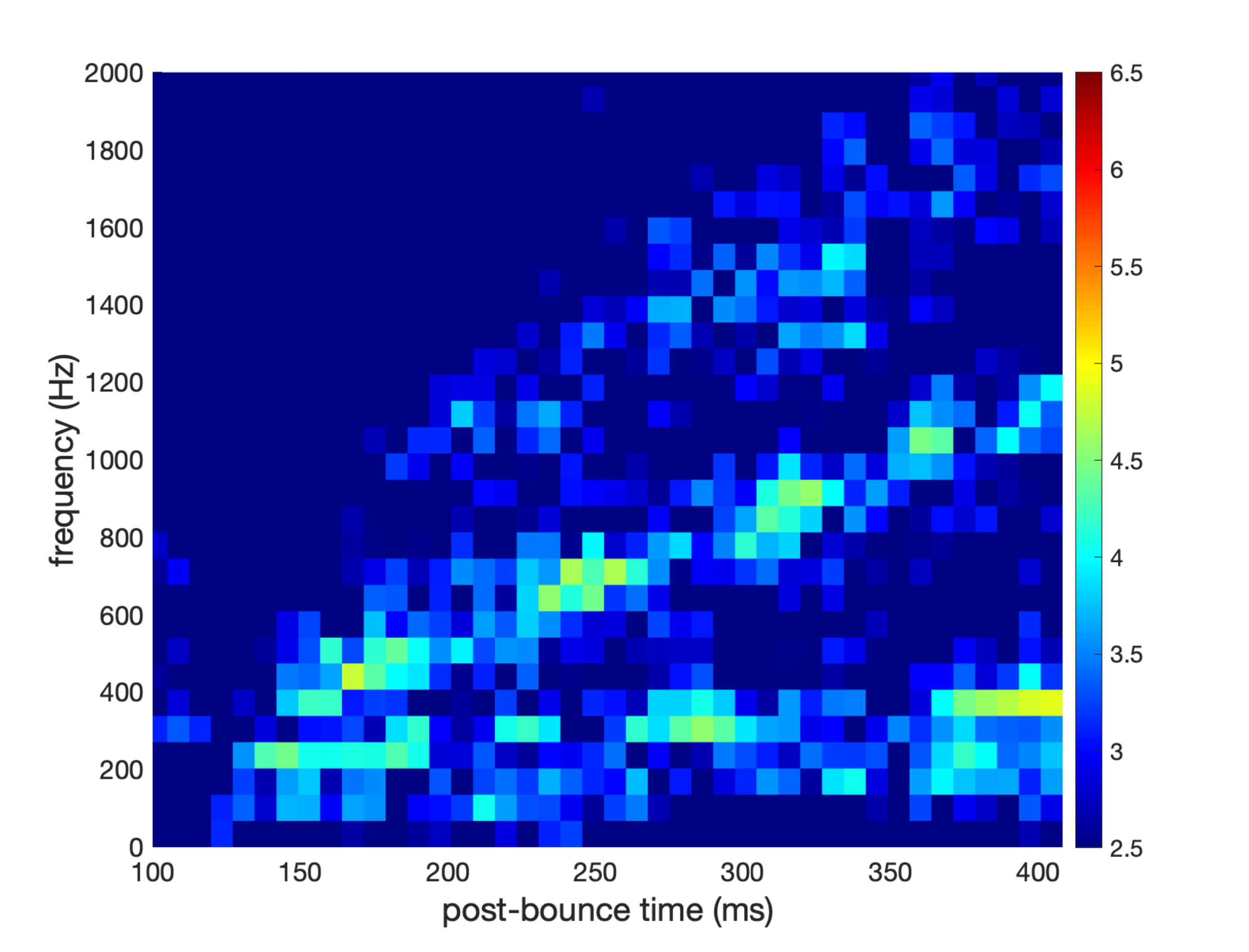}
\caption{Same as in Figure \ref{fig:heatmapfig} but for Region 3.}
\label{fig:heatmap3}
\end{figure}

\begin{figure}[ht!]
\centering
\includegraphics[width=80mm]{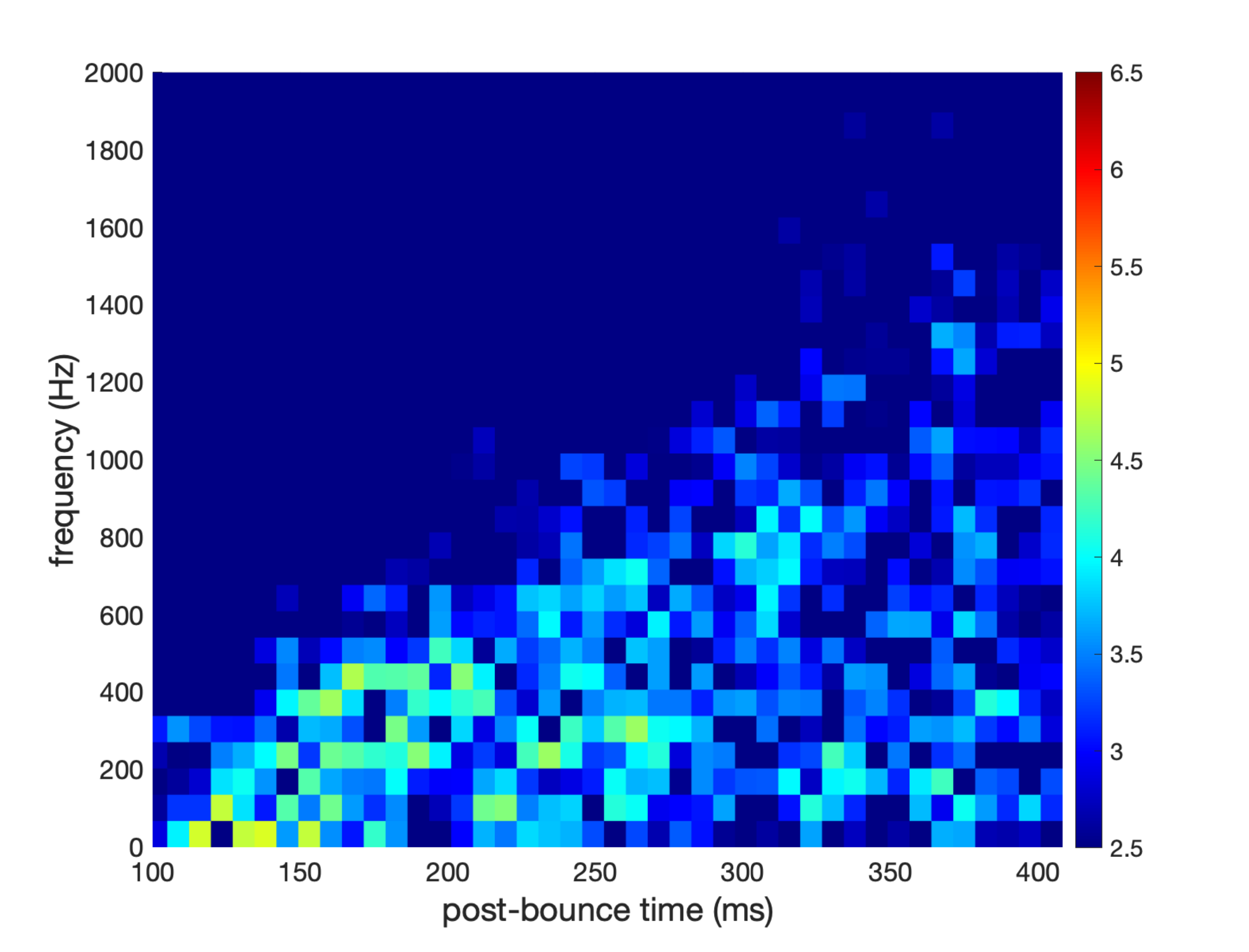}
\caption{Same as in Figure \ref{fig:heatmapfig} but for Region 4.}
\label{fig:heatmap4}
\end{figure}

\begin{figure}[ht!]
\centering
\includegraphics[width=80mm]{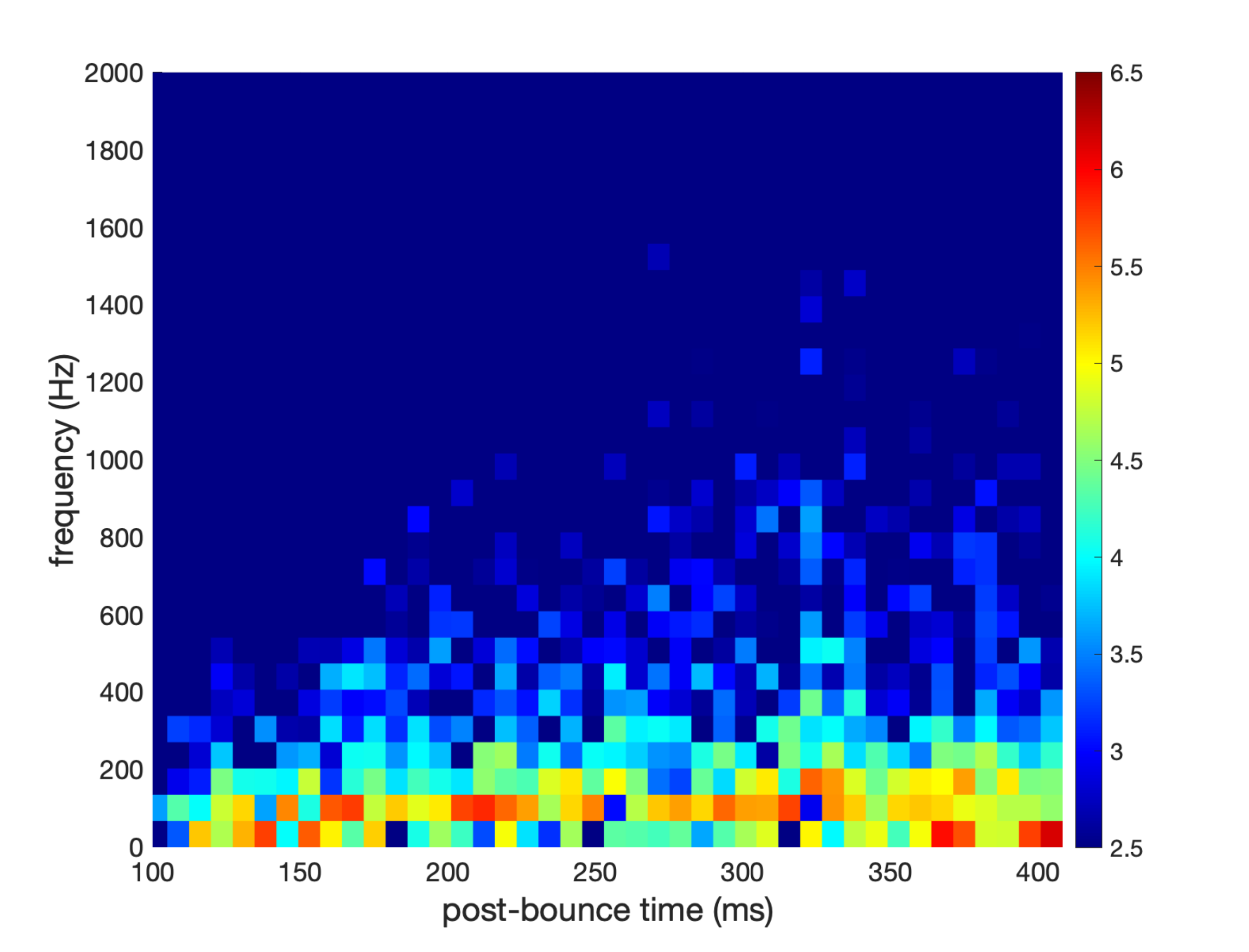}
\caption{Same as in Figure \ref{fig:heatmapfig} but for Region 5.}
\label{fig:heatmap5}
\end{figure}

\begin{figure}[ht!]
\centering
\includegraphics[width=80mm]{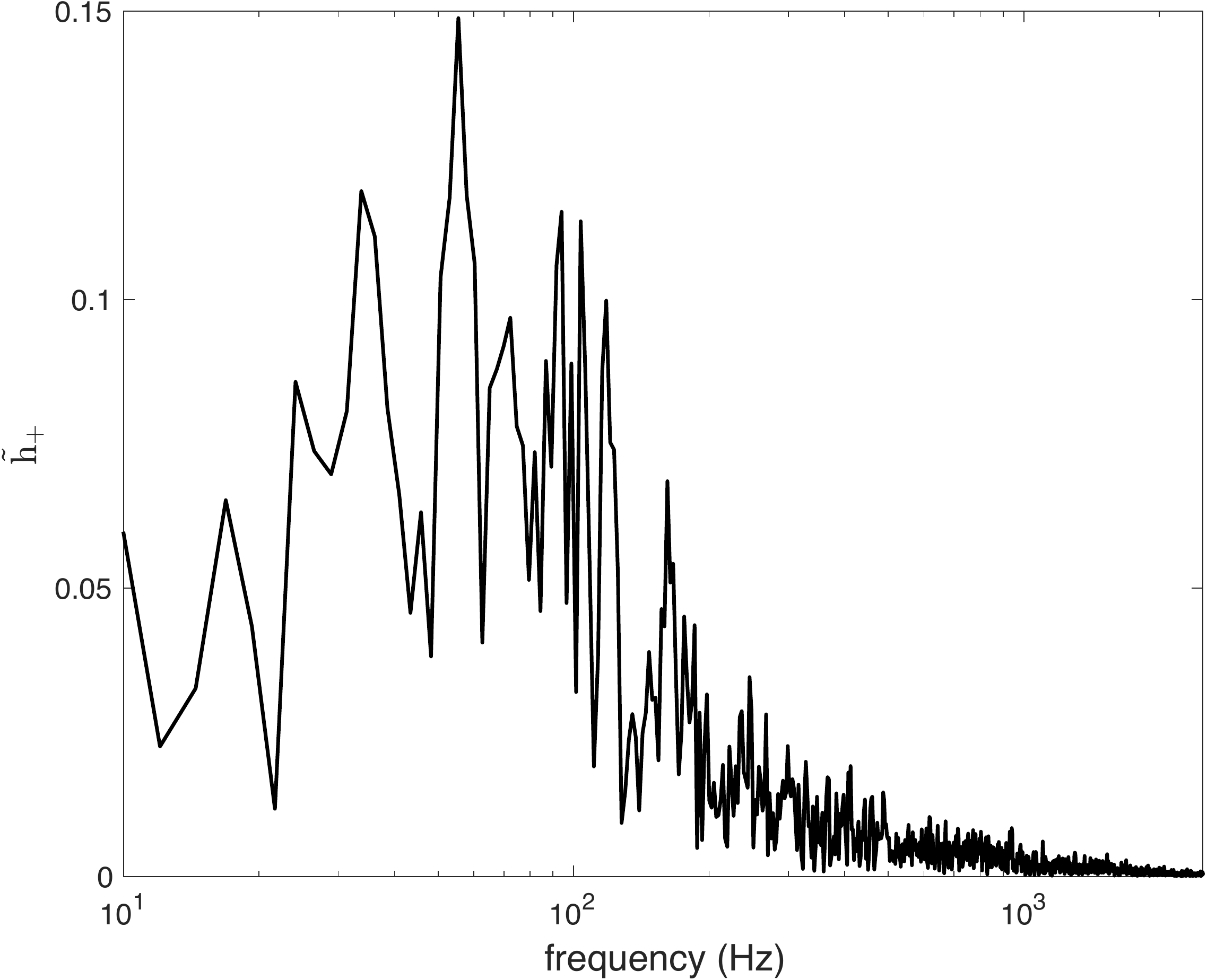}
\caption{The spectrum of gravitational wave emission from Region 5. The peak in the spectrum is well correlated with the expected emission frequency of the $\ell=1$ SASI mode. The spectrum also reflects contributions from (1) neutrino-driven convection, whose expected emission frequency begins at $\sim$20 Hz, with higher-frequency contributions expected, as well, due to the cascade of turbulent eddies to smaller spatial scales, expected and observed in our three-dimensional model, and (2) higher-order SASI modes -- in particular, the $\ell=2$ mode.}
\label{fig:region5spec}
\end{figure}

From Figure \ref{fig:rh+bylayers} and Figures \ref{fig:heatmap1} through \ref{fig:heatmap5}, the origin of the gravitational wave emission as a function of radial region and $t>100$ ms post bounce can be extracted. Figure \ref{fig:rh+bylayers} makes clear that the largest gravitational wave strains are generated in the convectively unstable layer: Region 1. Following the classification in Torres-Forn\'{e} et al. \cite{Torres-Forne19}, based on the nature of the restorative force in a region, we associate such emission with p-modes. Region 2, which is convectively stable, produces notable strains, which we associate with g-modes, resulting from convective overshoot from Region 1. Region 5 is the gain region and is clearly the source of our low-frequency -- i.e., frequencies below $\sim$200 Hz -- gravitational wave emission.
Gravitational wave emission in Region 5 results from mass motions induced by neutrino-driven convection and the SASI. We associate the emission in this region with p-modes. This breakdown of the gravitational wave emission by radial layer is born out in the heat maps for each of these layers, shown in Figures \ref{fig:heatmap1} through \ref{fig:heatmap5}. The high-frequency emission clearly stems from Regions 1 and 2, whereas the low-frequency emission, below $\sim$200 Hz, clearly stems from Region 5. Although of secondary importance here, it is worth noting that Regions 1 and 2 also emit gravitational radiation in the frequency range between 200 and 400 Hz. We attribute emission from these regions at these frequencies to accretion onto the proto-neutron star from above and the resultant excitation of additional g-modes within it. This emission is well correlated with the development of neutrino-driven convection and the SASI. As can be seen from Figures \ref{fig:heatmap1},  \ref{fig:heatmap2}, and \ref{fig:heatmap5}, emission at these frequencies from these regions begins after neutrino-driven convection and the SASI develop and is most pronounced near the end of our run, as explosion develops in the model.

Figure \ref{fig:region5spec} shows the spectrum of gravitational wave emission from the gain layer. To discern to the extent possible which features of the spectrum arise from neutrino-driven convection and which arise from the SASI, we compute the characteristic timescales associated with both and, in turn, their expected characteristic gravitational wave emission frequencies. We begin by estimating the convective overturn timescale in the gain layer, which is given by

\begin{equation}
\tau \sim \frac{2(R_{\rm Shock} - R_{\rm Gain})}{v_{\rm Convective}}.
\label{eq:convectiveoverturntimescale}
\end{equation}

\begin{figure}[ht!]
\centering
\includegraphics[width=80mm]{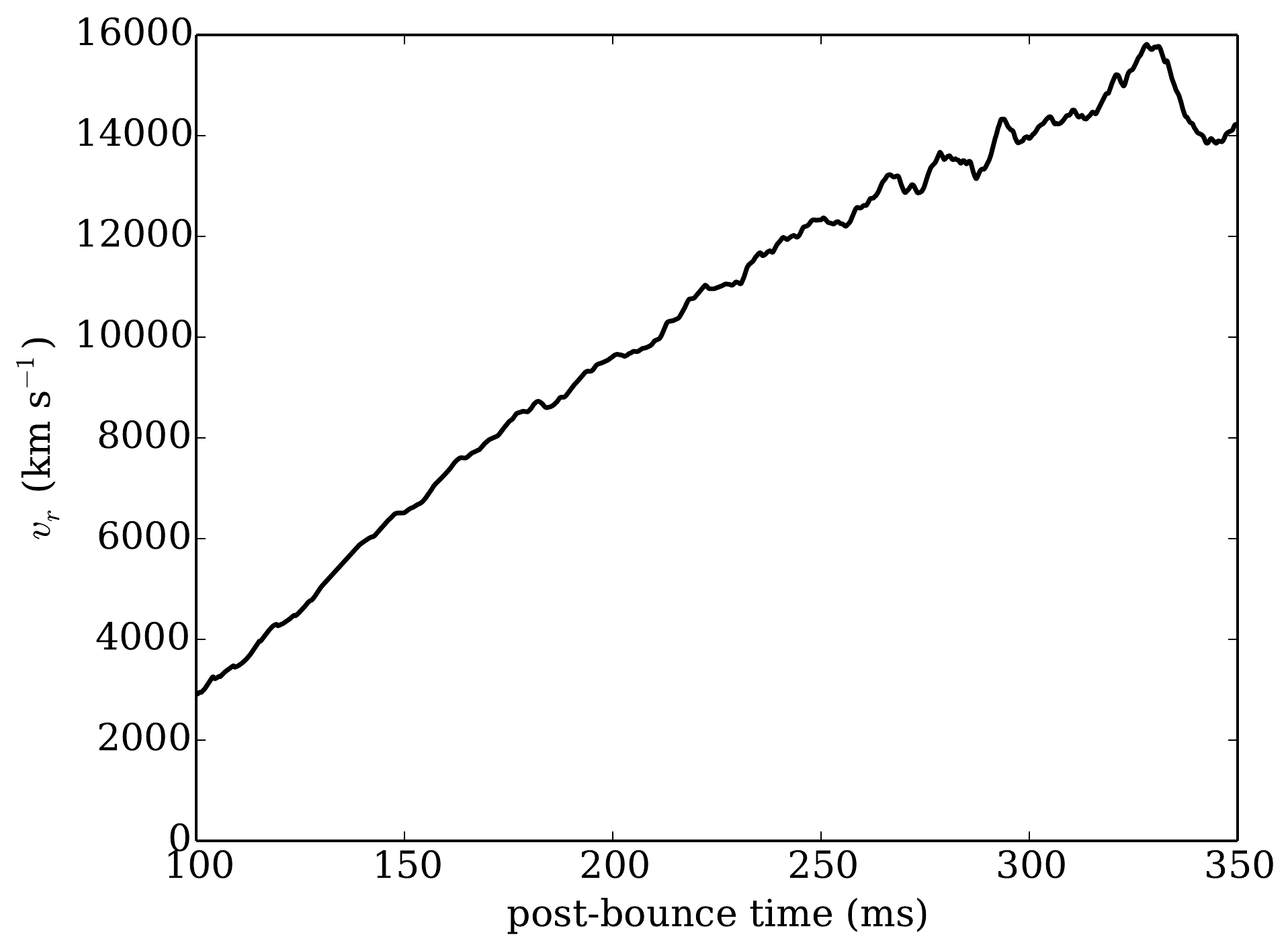}
\caption{A plot of the mean radial velocity in the gain region (Region 5) as a function of time after bounce. The mean radial velocity provides a measure of the 
convective overturn time scale in the gain region, which in turn provides a measure of the anticipated gravitational wave emission frequency from convection
in the region.}
\label{fig:meanradialvelocity}
\end{figure}

\begin{figure}[ht!]
\centering
\includegraphics[width=80mm]{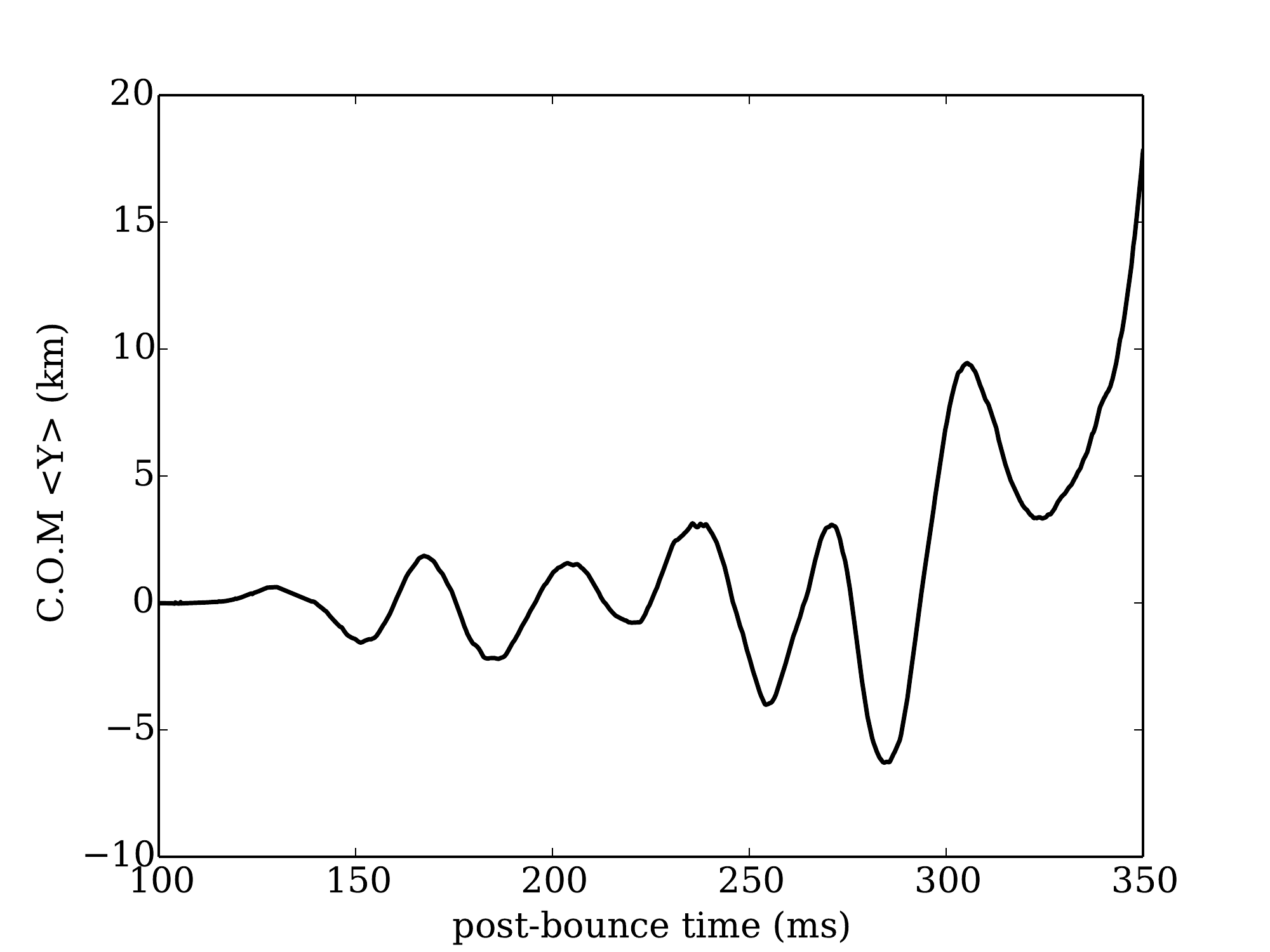}
\caption{A plot of the evolution of the center of mass of the cavity between the proto-neutron star surface and the shock, projected along the $y$-direction, as a function of time after bounce. Such evolution is a marker for the SASI. A steady oscillation of the center of mass, with a period of $\sim$35 ms, between $\sim$150 ms and $\sim$300 ms, at which time explosion develops in our model, is evident in the plot.}
\label{fig:COM-cavity}
\end{figure}

\noindent At 200 ms after bounce in our model, our angle-averaged shock radius $R_{\rm Shock}\approx$ 205 km and our angle-averaged gain radius $R_{\rm Gain}\approx $ 75 km. Looking at Figure \ref{fig:meanradialvelocity}, which shows the mean radial velocity in the gain region (Region 5) as a function of time after bounce, the mean radial velocity at this time is $\sim$6000 km s$^{-1}$. The convective overturn timescale is then $\sim$43 ms, which corresponds to an emission frequency $\sim$23 Hz. At 300 ms after bounce, $R_{\rm Shock}\approx $ 250 km, $R_{\rm Gain}\approx $ 56 km, and $v_{\rm Convective}\sim$ 7800 km s$^{-1}$, which gives $\tau\sim$50 ms and an emission frequency $\sim$20 Hz. Such emission frequency estimates will of course correspond to the lowest, or ``injection,'' frequencies. In our three-dimensional model, the cascade of large-scale convective eddies to smaller scales is expected, and with such a cascade, higher-frequency gravitational wave emission should also be expected.

Figure \ref{fig:COM-cavity} plots the evolution of the center of mass of the fluid within the cavity between the surface of the proto-neutron star and the shock, 
projected along the $y$-axis. The evolution of the center of mass is due to the SASI and a marker of its presence in our post-shock flow. Measuring from trough to trough or peak to peak, beginning at approximately 150 ms after bounce and continuing until 300 ms, after which explosion commences, we find that the SASI period remains remarkably steady, at $\sim$35 ms, corresponding to a SASI cycle frequency of $\sim$29 Hz. In every $l=1$ SASI cycle, two quadrupole deformation cycles result. Consequently, the SASI-induced gravitational wave emission from the $l=1$ mode is expected to occur at double the frequency -- i.e., at a frequency $\sim$58 Hz. This prediction agrees remarkably well with the location of the peak of the low-frequency spectrum we observe, shown in Figure \ref{fig:region5spec}. 

\begin{figure}[ht!]
\centering
\includegraphics[width=80mm]{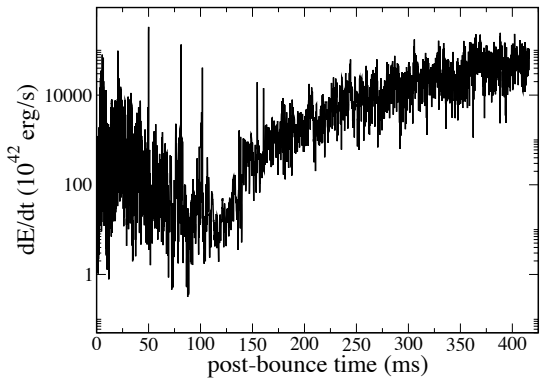}
\caption{The gravitational wave luminosity as a function of time after bounce. The time evolution of the luminosity follows the development of the instabilities in the stellar core that give rise to gravitational wave emission. The early rise and fall are associated with prompt convection. This is followed by a second rise in the gravitational wave luminosity after $\sim$100 ms after bounce, due to the development of aspherical mass motions in the gain layer from neutrino-driven convection and the SASI. The gravitational wave luminosity continues to rise in our model, through to the end of our run, as gravitational wave generation in the gain layer is joined by emission from the proto-neutron star, after $\sim$150 ms, due to emission from accretion onto it, and after $\sim$200 ms, due to Ledoux convection within it.}
\label{fig:GWluminfig}
\end{figure}

\begin{figure}[ht!]
\centering
\includegraphics[width=80mm]{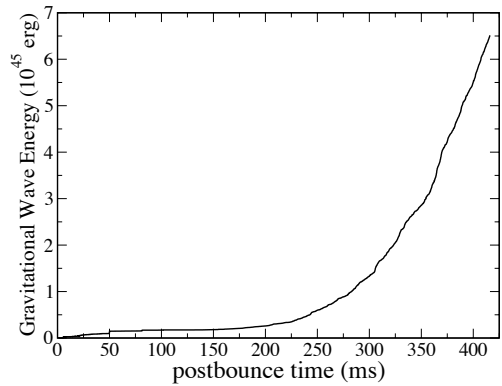}
\caption{The total gravitational wave energy emitted as a function of time after bounce. The significant rise in the gravitational wave energy production after $\sim$200 ms of post-bounce time results from the onset of Ledoux convection deep within the proto-neutron star, which persists for the remainder of the simulation.
The sharp rise is further evidence of the dominance of this phase of Ledoux convection in the proto-neutron star for the production of gravitational waves, despite the occurrence of several other sources -- prompt convection, neutrino-driven convection, the SASI, and accretion onto the proto-neutron star -- prior to its development.}
\label{fig:GWenergyfig}
\end{figure}

Figure \ref{fig:GWluminfig} shows the gravitational wave luminosity as a function of time after stellar core bounce. 
We attribute the initial rise and fall of the luminosity between the beginning of our run and $\sim$100 ms after bounce to the development of prompt convection in the proto-neutron star and the ensuing three-dimensional flows it induces in the region.
The production of gravitational wave energy then rises after $\sim$100 ms, given the development of aspherical mass motions in the gain layer due to neutrino-driven convection and the SASI. This period of gravitational wave energy production is then followed by a final period of production, beginning at $\sim$150 ms after bounce, during which time gravitational wave emission emanates from the proto-neutron star, initially due to accretion onto it from above, and later due also and predominantly to Ledoux convection within it. During this last period of gravitational wave emission, the gravitational wave amplitude continues to rise, as shown in Figure \ref{fig:rh+fig}, and the gravitational wave energy luminosity continues to rise, as well, though at a decreasing rate toward the end of our run.

Figure \ref{fig:GWenergyfig} gives the total energy emitted in gravitational radiation as a function of time after bounce. Despite multiple sources contributing to gravitational radiation emission before $\sim$200 ms after bounce, which includes prompt convection (to a small degree), neutrino-driven convection, the SASI, and accretion onto the proto-neutron star from above, the gravitational wave energy emitted remains low until the second and lasting phase of proto-neutron star Ledoux convection begins.

\begin{figure}[ht!]
\centering
\includegraphics[width=80mm]{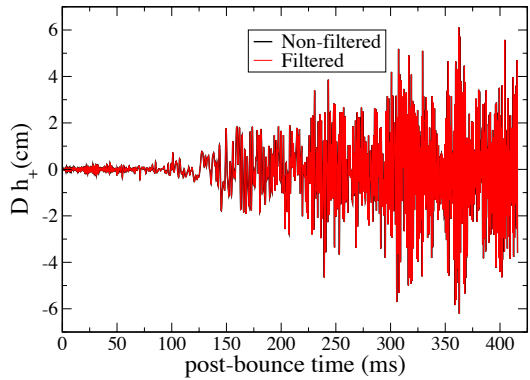}
\caption{A comparison of the filtered and unfiltered gravitational wave strains as a function of time after stellar core bounce, for the $+$ polarization. The filtering removes the contributions to the early gravitational wave strain from the region between the initial Ledoux unstable region and the shock, due to numerically-induced entropy fluctuations, which can give rise to gravitational wave emission. It also removes any gravitational wave strain from regions ahead of the shock induced by the instantaneous transmission of aspherical gravity within the core 
to this region given our use of a gravitational potential. An aspherical potential can induce aspherical flows in this otherwise spherically collapsing region.}
\label{fig:rh+unfiltered}
\end{figure}

\subsection{Filtering}

During the early course of our run ($t<70$ ms of post-bounce evolution), we have identified two sources of numerically-induced, though small-amplitude, gravitational wave emission: (1) We obtained radial fluctuations in the entropy profile in the post-shock region induced by our hydrodynamics method (a well-known artifact of the PPM method \cite{DONAT199642,JIN1996373}). When convolved with the angular dependence found in the formulae to compute the gravitational wave strain [e.g., see Equation (\ref{eq:n2m-integration})], these entropy ``wiggles'' resulted in a numerically-induced gravitational wave strain. 
Given the numerical origin of these entropy fluctuations, to the extent possible we filtered out any contributions by them to the gravitational wave strain. This was accomplished by truncating the radial integration in Equation (\ref{eq:n2m-integration}) at the outer radius of the initial Ledoux unstable region shown in Figure \ref{fig:BVfig}. This removed any gravitational wave emission from numerically-induced entropy fluctuations above this region. At early times it is possible to have limited overlap between the Ledoux unstable region and the region in which the initial entropy fluctuations occur. In this case, it is not possible to separate physical from numerically-induced signals fully. The radius at which the truncation was imposed by our filtering method was delineated as a function of time in our run, and in turn used in the post-processing of our gravitational wave strains, until the need for filtering disappeared. Once filtering ceased at $70$ ms after bounce, the integral in Equation (\ref{eq:n2m-integration}) was carried out over our entire numerical domain. 
(2) Given our use of a gravitational potential, which can carry information from one region in the core to another instantaneously, we observe mildly aspherical flows ahead of the shock in an otherwise spherically symmetric infalling fluid. These flows result from the aspherical potential given the aspherical flows deep within the core, from prompt convection.
Figure \ref{fig:rh+unfiltered} compares the filtered and unfiltered gravitational wave strains for the $+$ polarization, as a function of post-bounce time. There is no appreciable difference between the strains as a result of filtering.

\begin{figure}[ht!]
\centering
\includegraphics[width=80mm]{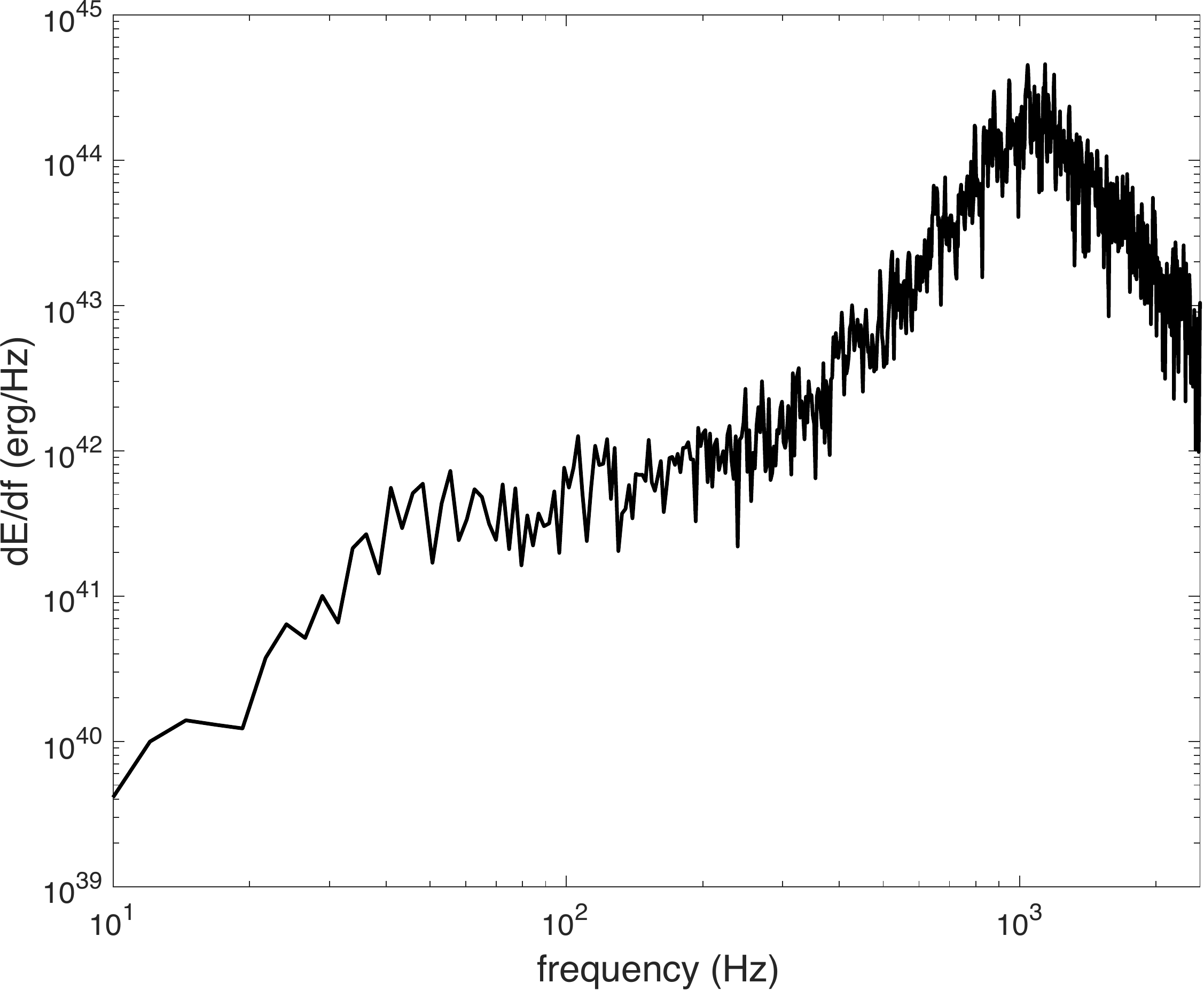}
\caption{The gravitational wave energy spectrum, computed at the end of our simulation. Most of the gravitational wave energy is emitted at frequencies $\sim$1 kHz, whose origin is persistent Ledoux convection in the proto-neutron star driven by continued deleptonization during the course of our simulation. As we move to lower frequencies, the spectrum decreases quickly until $\sim$400 Hz, at which point its rate of decline slows considerably until $\sim$40 Hz, at which point it again drops quickly. Gravitational emission between $\sim$40 Hz (and below) and $\sim$400 Hz has its origins in the mass motions in the gain layer due to neutrino-driven convection and the SASI and to the resultant aspherical accretion onto the proto-neutron star.}
\label{fig:dedffig}
\end{figure}

\begin{figure}[ht!]
\centering
\includegraphics[width=80mm]{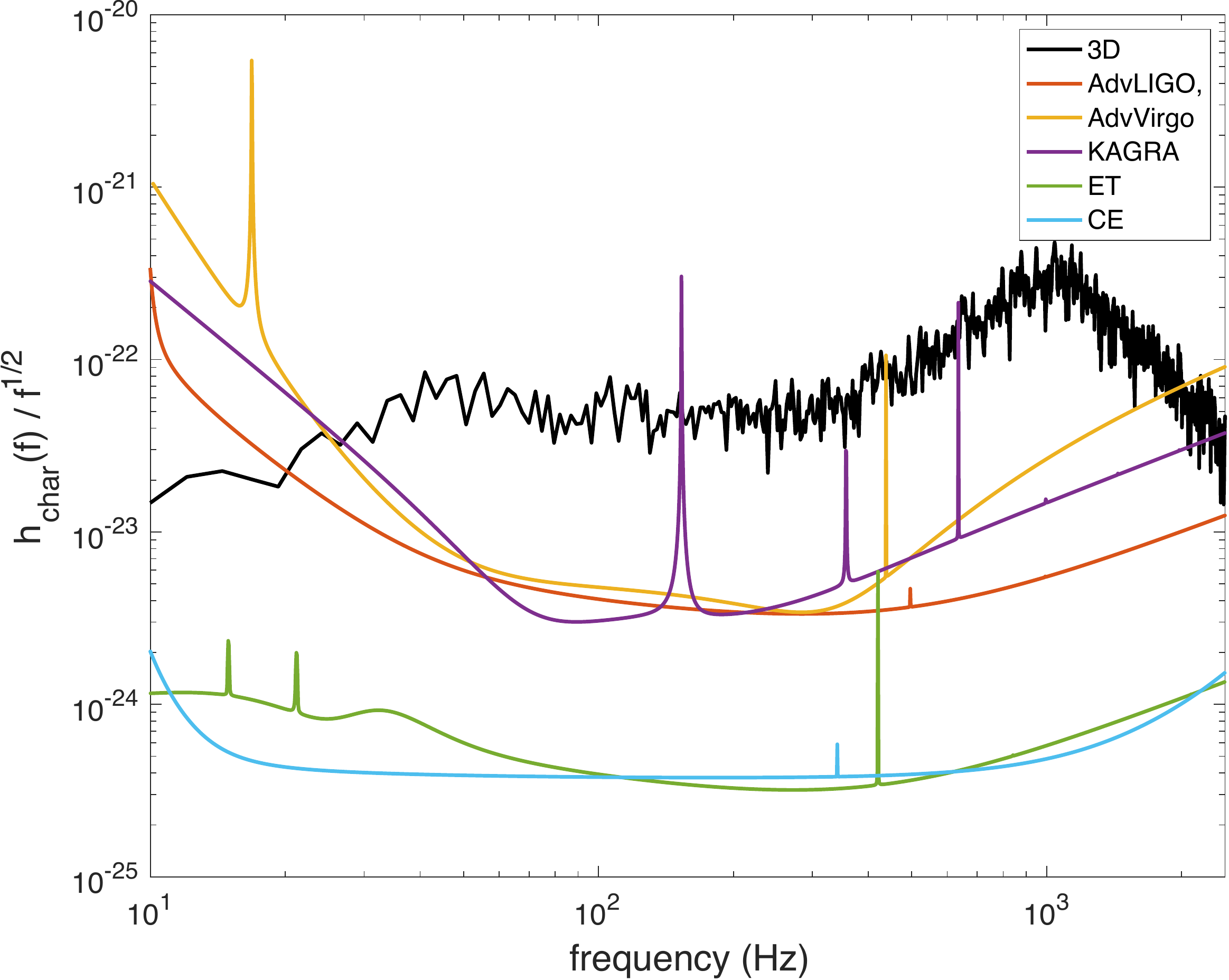}
\caption{The characteristic gravitational wave strain plotted as a function of frequency for a supernova at 10 kpc. 
Also shown are the sensitivity curves for the current-generation gravitational wave detectors Advanced LIGO, Advanced VIRGO, and KAGRA, and the next-generation Cosmic Explorer and Einstein Telescope (D configuration).
Detection of the gravitational wave signal from a core collapse supernova across the full spectrum of emission that would bring information about both neutrino-driven convection/SASI activity and proto-neutron star Ledoux convection will require sensitivities as low as $\sim3 \times 10^{-22}$ at frequencies above $\sim$20 Hz, which, except for the lowest frequencies between $\sim$20 Hz and $\sim$40 Hz, is satisfied by all of the detectors included here.}
\label{fig:hcharfig}
\end{figure}

\subsection{Spectral Analysis}

The gravitational wave energy spectrum computed at the end of our run is shown in Figure \ref{fig:dedffig}. The spectrum peaks just above 1 kHz. It is clear the gravitational wave energy emission is dominated by high-frequency emission. We associate this part of the spectrum with long-lived Ledoux convection deep within the proto-neutron star. The spectrum also features two breaks. As we move from the peak frequency to lower frequencies, the spectrum drops precipitously until we reach a frequency of $\sim$400 Hz, at which point the spectrum levels off. Between $\sim$400 Hz and $\sim$40 Hz, the spectrum varies more slowly with frequency. Below 
$\sim$40 Hz, it again drops off rapidly. The spectrum in the frequency range between $\sim$40 Hz (and below) and $\sim$400 Hz is sustained by neutrino-driven convection, the SASI, and accretion of convection- and SASI-induced aspherical flows onto the proto-neutron star.

Figure \ref{fig:hcharfig} shows the characteristic gravitational wave strain as a function of frequency, for a supernova at 10 kpc. Also shown are the sensitivity curves for the current-generation gravitational wave detectors Advanced LIGO \cite{Abbott2018}, Advanced VIRGO \cite{Abbott2018}, and KAGRA \cite{Abbott2018}, and the next-generation Cosmic Explorer \cite{Abbott2017sensitivity} and Einstein Telescope (D configuration) \cite{Punturo2010,Hild2011}.
Sufficient sensitivity to strains down to $\sim$few $\times 10^{-23}$, at frequencies between $\sim$20 Hz up through $\sim$1 kHz, would provide much of the core collapse supernova gravitational wave emission spectrum, from the low-frequency emission associated with mass motions in the gain layer, linked to neutrino-driven convection and the SASI, to the high-frequency emission associated with proto-neutron star Ledoux convection. Such sensitivity for a Galactic event (10 kpc) is exhibited by all of the gravitational wave detectors included here.

\begin{figure}[ht!]
\centering
\includegraphics[width=80mm]{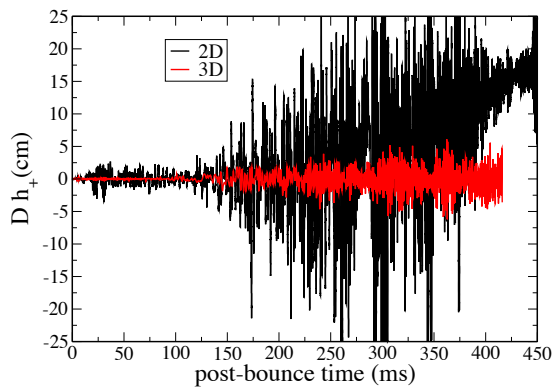}
\caption{The gravitational wave strain for the $+$ polarization plotted as a function of post-bounce time over the course of our simulation, for both the two-dimensional (black) and the three-dimensional (red) models. In the two-dimensional case, the view is along the $x$-axis, whereas in the three-dimensional case, it is along the $z$-axis. In axisymmetry (about the $z$-axis), the strain viewed along the $z$-axis is zero. The strains plotted here are differentiated by two key factors. The strain amplitude  for the three-dimensional case is significantly smaller, and the time to explosion in this case (as marked by the offset of the gravitational wave strain in the two-dimensional case) is significantly delayed.
}
\label{fig:rh+2Dv3Dfig}
\end{figure}

\begin{figure}[ht!]
\centering
\includegraphics[width=80mm]{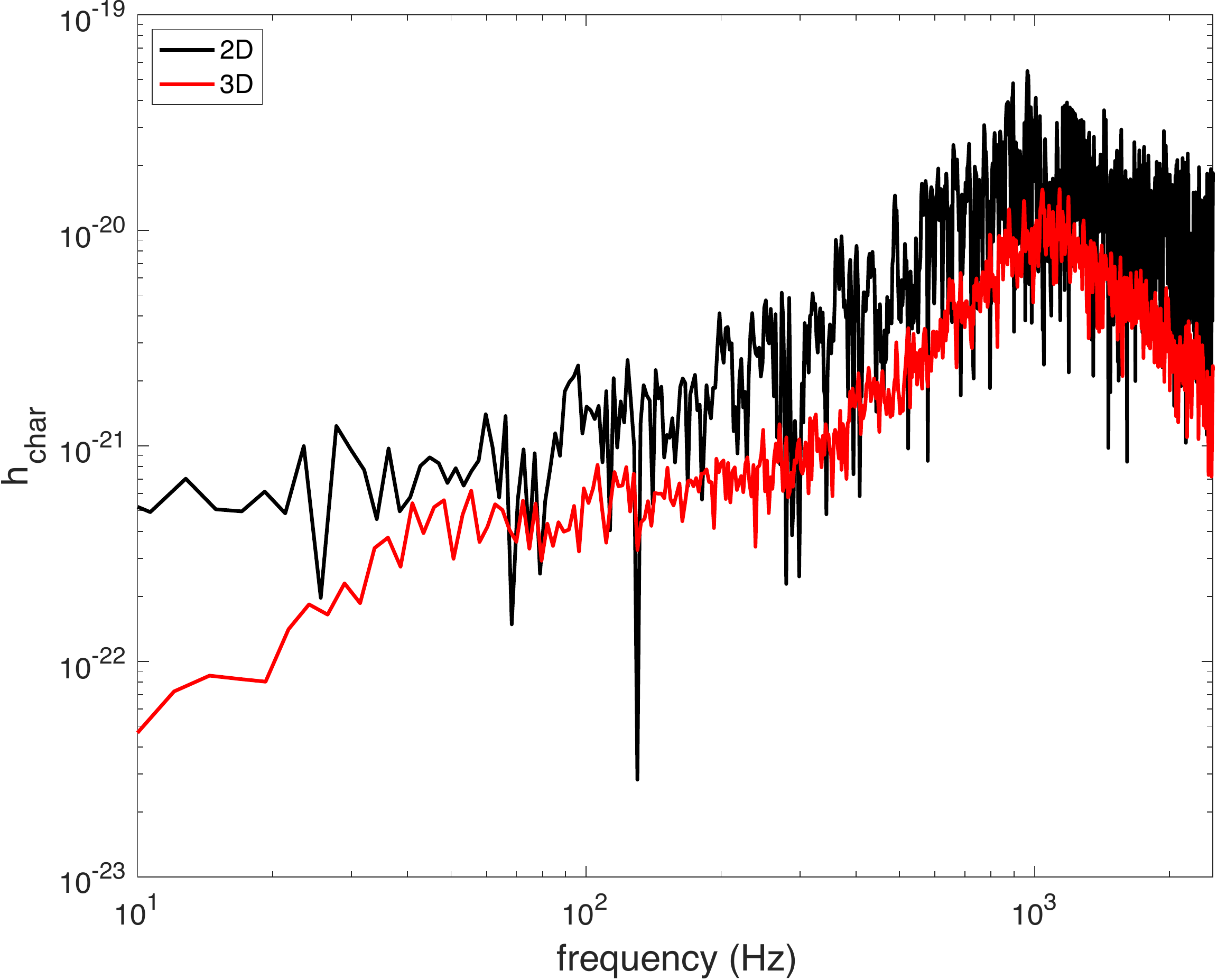}
\caption{The characteristic gravitational wave strain plotted as a function of frequency, for both our two-dimensional and our three-dimensional models. The peak frequency is somewhat higher in the three-dimensional case, but the gravitational wave energy across the spectrum is significantly less. At the lowest frequencies, the characteristic strains for the two cases diverge. This is the result of the well-known accumulation of kinetic energy at larger spatial scales in two dimensions, resulting in greater gravitational wave energy at lower frequencies.}
\label{fig:hchar2Dv3Dfig}
\end{figure}

\begin{figure}[ht!]
\centering
\includegraphics[width=80mm]{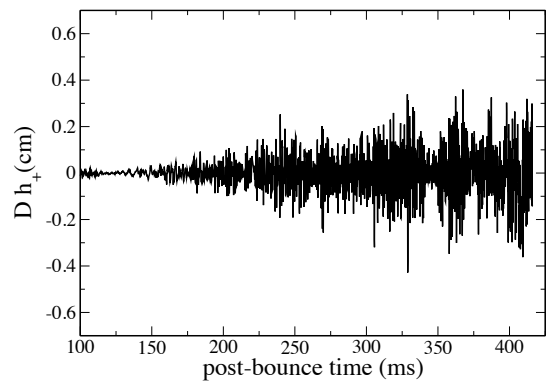}
\caption{The $+$ polarization gravitational wave strain for the innermost region in the proto-neutron star in our model, below the region of sustained Ledoux convection dominating our gravitational wave emission. The strain amplitudes from this region are a small fraction of the amplitudes from the region above it, indicating little contribution to the gravitational wave emission in this model from convective undershoot.}
\label{fig:rh+innermost}
\end{figure}

\subsection{Comparison of Two-Dimensional and Three-Dimensional Gravitational Wave Emission}

Figures \ref{fig:rh+2Dv3Dfig} and \ref{fig:hchar2Dv3Dfig} show the differences between the gravitational wave strain and the dimensionless characteristic strain based on our two- and three-dimensional models (in both cases, the supernova is presumed to be at a distance of 10 kpc). Focusing first on the gravitational wave strain, two key differences stand out: (1) The gravitational wave amplitudes are significantly lower in three dimensions. (2) Explosion is significantly delayed in three dimensions. The overall rise in the gravitational wave amplitude after $\sim$225 ms in our two-dimensional model is evidence of the onset of explosion. Indeed, based on the angle-averaged shock trajectory and evolution of the diagnostic energy, explosion in the two-dimensional model sets in at this time \cite{Lentz15}. 

Regarding the dimensionless characteristic strains: The two- and three-dimensional strains are both dominated by high-frequency (proto-neutron star) emission. Across the spectrum, the amplitude of the characteristic strain is on average significantly lower in the three-dimensional case, although there are large variations in the strain in the two-dimensional case that make comparison of the amplitudes difficult. Below $\sim$40 Hz, the characteristic strains for the two- and three-dimensional cases diverge. In two dimensions, symmetry constraints promote the growth of long-wavelength, low-frequency mass motions. No such constraints are present in three dimensions, and the behavior of the characteristic strain simply reflects the character of neutrino-driven convection in three dimensions, as we have shown.

\begin{figure}[ht!]
\centering
\includegraphics[width=80mm]{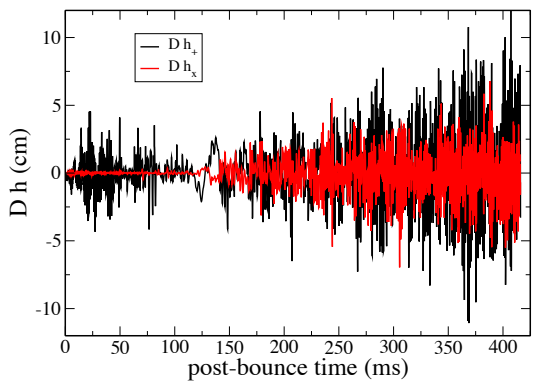}
\caption{Plot of the strain amplitude for both the $+$ and the $\times$ polarizations, viewed along the $x$-axis. The impact of the constant-$\mu$ grid on the magnitude of the $+$-polarization strain is apparent, where now the expectation that the strain amplitudes for the $+$ and $\times$ polarizations be comparable due to the lack of a preferred physical direction in our model is no longer satisfied.}
\label{fig:rh+rhx_xaxis}
\end{figure}

\begin{figure}[ht!]
\centering
\includegraphics[width=80mm]{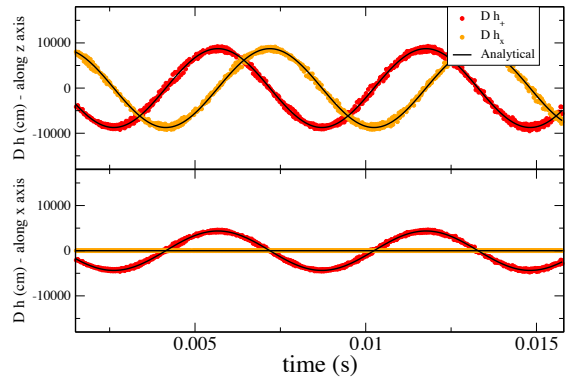}
\caption{Gravitational wave predictions for both polarizations in the case of a test using orbiting neutron stars for which an analytical solution is available. In this case, the binary is set to orbit in the $xy$-plane, and the gravitational wave emission is observed along both the $x$-axis and the $z$-axis.}
\label{fig:BNSTest_xy}
\end{figure}

\begin{figure}[ht!]
\centering
\includegraphics[width=80mm]{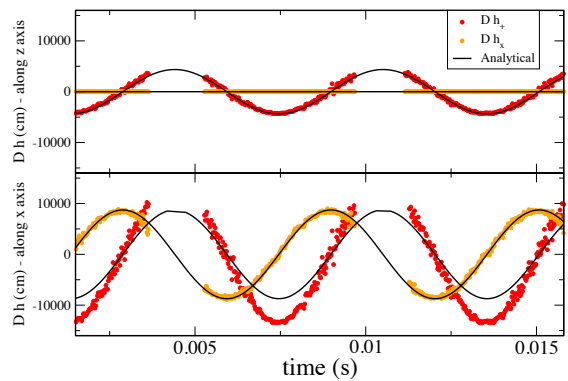}
\caption{Same as in Figure \ref{fig:BNSTest_xy} but now the neutron stars are made to orbit in the $yz$-plane. Points are excluded when the binary crosses the $z$-axis, where the numerical results fluctuate and are no longer illustrative for our present purpose. When viewed along the $z$-axis, the strains agree well with the analytical result except when a pole crossing occurs. On the other hand, when viewed along the $x$-axis, the $+$-polarization strain is consistently larger than the analytical result, whereas the $\times$-polarization strain continues to agree well with it.}
\label{fig:BNSTest_yz}
\end{figure}

\begin{figure}[ht!]
\centering
\includegraphics[width=80mm]{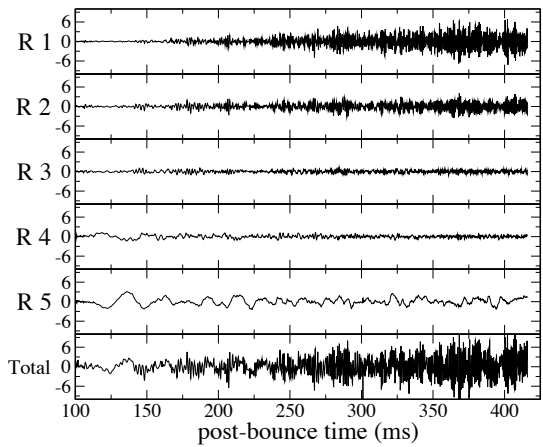}
\caption{Same as in Figure \ref{fig:rh+bylayers} but viewed along the $x$-axis. Despite the numerical effects associated with the constant-$\mu$ grid, our conclusions regarding the dominant layers of low- and high-frequency gravitational wave emission are unchanged.}
\label{fig:rh+bylayer_xaxis}
\end{figure}

\subsection{Other, Numerical Considerations}
\label{sec:numericalconsiderations}

For times $t>100$ ms after bounce, Figure \ref{fig:rh+innermost} shows the gravitational wave strain (for the $+$ polarization) from the region below the Ledoux unstable region in our run. We show this for two reasons: (1) To demonstrate that the stable region below the Ledoux unstable region is a negligible source of gravitational radiation. (2) To quantify to what extent our strain predictions are impacted by the fact that we impose spherical symmetry below $\sim$8 km. 
With regard to (1), the maximum amplitudes observed in this region over the course of our run are $\sim$0.3 cm, whereas the amplitudes in the convective layer above reach maxima that are of $\sim$6 cm. Moreover, that any strain is associated with the region below the convective layer results in part from the definition of the boundary between it and the convective layer, which in our analysis is done through spherical averaging. A small shift of the boundary inward could result in the elimination of this contribution to the strain.
With regard to (2), our imposition of spherical symmetry within a very small volume at the center of our model likely had some impact on our determination of the gravitational wave radiation emitted from the region directly above it (i.e., the region in question: the region below the convective layer), but given the disparate magnitude of the strain amplitudes in the region in question relative to the convective region above it and given the fundamentally different nature of the gravitational wave emission from these two adjacent regions, one being convectively stable and one being convectively unstable, we think it unlikely that our limited imposition of spherical symmetry in our simulation would have fundamentally altered the results we present here and the conclusions we have drawn from them. However, definitive conclusions will have to wait on more advanced simulations.

In Figure \ref{fig:rh+rhx_xaxis} we plot the gravitational wave strains in our model, for both polarizations, but now viewed along the $x$-axis -- ie., at a viewing angle of $\theta=\pi/2$ (and, as before, $\phi=0$). By doing so, we are able to see the impact of our use of a constant-$\mu$ grid, with lower angular resolution at $\theta=0$ and higher angular resolution at $\theta =\pi/2$, on our gravitational wave strain predictions. In particular, we can see that the amplitude of the strain for the $+$ polarization is affected by the change in viewing angle, whereas the amplitude of the strain for the $\times$ polarization is not. The ratio of the two is a factor $\sim$2 for a viewing angle along the $x$-axis, whereas in Figure \ref{fig:rh+rhx_zaxis}, corresponding to a viewing angle along the $z$-axis, they are comparable. This is an artifact of our constant-$\mu$ grid. To prove this, we include here the results of a test of our code that has an analytical solution. We consider the gravitational waves emitted by binary neutron stars in Keplerian orbit about one another. We consider two cases: (1) binary orbit in the $xy$-plane and (2) binary orbit in the $yz$-plane. The binary orbit in the $xy$-plane should not be affected by our use of a constant-$\mu$ grid. In Figure \ref{fig:BNSTest_xy}, we see that the predictions made by our code for both polarizations and both viewing angles are in close agreement with the analytical solution. On the other hand, Figure \ref{fig:BNSTest_yz} shows that the $+$-polarization strain is consistently larger than the analytical solution when viewed along the $x$-axis, whereas the $\times$-polarization strain continues to agree with the analytical result except when the binary orbit crosses the $z$-axis. Moreover, with the exception of the pole crossing, both strains agree well with the analytical solution when the strains are viewed along the $z$-axis. From the point of view of mass distribution, this test represents an extreme not sampled in our model. The results shown in Figures \ref{fig:BNSTest_xy} and \ref{fig:BNSTest_yz} together demonstrate that our predictions for gravitational wave emission for our model are robust for a viewing angle along the $z$-axis, as chosen in the analysis presented in this paper.
Finally, in Figure \ref{fig:rh+bylayer_xaxis} we plot the $+$-polarization gravitational wave strains for Regions 1--5, as in Figure \ref{fig:rh+bylayers}, but now for a viewing angle along the $x$-axis. Our conclusion that our gravitational wave emission is dominated by emission from Region 1, the region of Ledoux convection, is unaltered, as is our conclusion that the dominant low-frequency source is Region 5.

\section{Summary, Discussion, and Conclusions}

We have performed an analysis of the gravitational wave emission based on the output data of our first three-dimensional core collapse supernova simulation performed with the \chimera code \cite{Lentz15}. We have herein discussed the temporal evolution -- specifically, the gravitational wave strain for both polarizations, peak frequency evolution, gravitational wave luminosity, and total emitted gravitational wave energy -- with time. We have also provided the gravitational wave energy spectrum and characteristic strain as a function of frequency, both computed at the end of our run. We also documented the sources of the gravitational wave emissions in our model. The results presented here differ significantly from those presented in Yakunin et al. \cite{Yakunin2017} and replace them. The associated gravitational wave data, available online (http://www.phys.utk.edu/smc), were replaced on March 7, 2019. We have updated the data, on May 23, 2020, with the data used for the analysis presented here.

The gravitational wave emission occurs in multiple phases. The first phase, which is fairly quiescent, during which time gravitational wave emission in either polarization is not significant, results from the development of prompt convection in the proto-neutron star shortly after bounce, and lasts $\sim$100 ms. The core is Ledoux unstable for only $\sim$30 ms after bounce, then becomes Ledoux stable as the convection stabilizes the entropy and lepton gradients. But the gravitational-wave-producing flows resulting from the initial convective instability persist. The second phase of gravitational wave emission, beginning after $\sim$100 ms after bounce and lasting until $\sim$150 ms after bounce results from the development of aspherical mass motions in the gain layer induced by neutrino-driven convection and the SASI. This phase is particularly evident in the sharp rise in the gravitational wave luminosity during this time frame. The low-frequency emission from these sources persists throughout the remainder of our run, evident in the heat map below frequencies $\sim$200 Hz. After $\sim$150 ms after bounce, a third phase of gravitational wave emission begins when emission from the above sources is joined by emission from the proto-neutron star due to (1) aspherical accretion onto it, which gives rise to intermediate-frequency emission in the range $\sim$400--600 Hz, and (2) a second phase of Ledoux convection, deep in the interior of the proto-neutron star, which gives to high-frequency gravitational wave emission. Both persist to the end of our run. The latter dominates the gravitational wave emission in our model. Unlike the first phase of proto-neutron star Ledoux convection, the second phase persists due to the maintenance of unstable lepton gradients given continued neutrino diffusion out of the core.
The gravitational wave amplitude in this third phase of emission grows with time. The peak frequency from Ledoux convection emission begins at $\sim$600 Hz at the start of convection and rises to $\sim$1200 Hz by the end of our run, as the proto-neutron star deleptonizes, its central density increases, and its radius declines. The gravitational wave luminosity associated with this final phase of emission continues to rise through to the end of our run, as well, and the total gravitational wave energy emitted rises dramatically after $\sim$200 ms after bounce, once significant Ledoux-convection-induced gravitational wave emission begins.

The spectrum of gravitational wave energy emitted peaks at a frequency just above 1 kHz. The emission at high frequency clearly dominates the total gravitational wave energy produced. We associate this part of the spectrum with the last phase of proto-neutron star convection. As we move to lower frequencies, the spectrum drops precipitously, between 1 kHz and $\sim$400 Hz, and then exhibits a more gradual decline down to $\sim$40 Hz, followed by a second drop off. We associate the region of the spectrum between $\sim$400 Hz and $\sim$40 Hz (and below) with the mass motions in the gain layer induced by neutrino-driven convection and the SASI and with gravitational wave emission from the proto-neutron star due to aspherical accretion onto it. Detectors with sensitivities down to $\sim$few $\times 10^{-23}$ at frequencies $\sim$20 Hz will be able to capture much of the spectrum of gravitational wave emission we have detailed here, between $\sim$20 Hz and $\sim1$ kHZ, and, with it, invaluable information about the underlying sources of emission in the supernova. Fortunately, for the event presumed here -- a core collapse supernova at a distance of 10 kpc -- such sensitivity will be provided by current-generation detectors (Advanced LIGO, Advanced VIRGO, and KAGRA).

Based on our model, we arrive at the same conclusion as that drawn by Andresen et al. \cite{Andresen2017} in the context of their models that the gravitational wave emission is dominated by late-time, long-lived Ledoux convection in the proto-neutron star -- i.e., that the gravitational wave energy produced stems largely from the fluid dynamics {\em within} the proto-neutron star, not from perturbations of the proto-neutron star by fluid dynamics above it. However, in our model, the dominant emission stems from the convective region itself, rather than from the convective overshoot layer above it, as was the case for the models they considered. 
Following the analysis of the characteristics of gravitational wave modes performed by Torres-Forn\'{e} et al. \cite{Torres-Forne19}, we associate the dominant emission in our model with p-modes. 
We do obtain gravitational wave emission in the convectively-stable overshoot layer, as seen by our isolation of the contribution from this layer to the gravitational wave strain. This emission, instead, is naturally interpreted as g-mode emission. Thus, our results suggest that the high-frequency gravitational wave emission from within the proto-neutron star in our model is hybrid emission from both p- and g-modes, though dominated by the former. Finally, in our model we do not obtain any gravitational wave emission of note in the surface layer of the proto-neutron star. Our conclusions and the conclusions of Andresen et al. therefore differ from those drawn by O'Connor and Couch \cite{O'Connor2018}, Radice et al. \cite{Radice2019}, and Powell and Mueller \cite{Powell2019} in the context of their models. They conclude that even in the three-dimensional case gravitational wave emission from the proto-neutron star, from its surface layers, is still excited by accretion funnels generated in the region between the proto-neutron star surface and the shock, due to the mass motions induced by neutrino-driven convection and the SASI.

In the model we consider here, we come to a different conclusion than Kuroda et al. \cite{Kuroda2016a} -- again, made in the context of their models -- with regard to the source of low-frequency gravitational wave emission. We find that emission of gravitational waves at frequencies below $\sim$200 Hz stems from mass motions in the gain layer and not from within the proto-neutron star due to low-frequency modulation of the accretion flow onto it. In the context of their models, Andresen et al. \cite{Andresen2017} conclude that low-frequency gravitational radiation is emitted from the gain layer, as well, but they conclude that layers below the gain layer, including the deep convective layer, contribute, too. They conclude, as do Kuroda et al., that low-frequency emission from layers below the gain layer results from SASI-induced, low-frequency, accretion-flow modulation.

Finally, Powell and Mueller \cite{Powell2019} point out that gravitational wave emission in their models is greatest after the onset of explosion and emphasize the need to push supernova models sufficiently far post bounce to capture the full gravitational wave signature. This was emphasized by Yakunin et al. \cite{Yakunin10,Yakunin15} as well. It is not simply a matter of running sufficiently long to capture what might potentially be the strongest phase of gravitational wave emission. It is a matter also of capturing accurately the phases of gravitational wave emission prior to explosion. The model we present here covered approximately the first half second of post-bounce evolution. In this particular model, explosion (as defined by both the angle-averaged shock trajectory and the diagnostic energy) begins at $\sim$300 ms after bounce \cite{Lentz15}. Thus, our analysis spans sufficient postbounce time to cover the majority of the explosion epoch, though our strain amplitudes continue to grow and the total gravitational wave energy emitted continues to increase at the end of our run. However, our gravitational wave luminosity, though still increasing at the end of our run, as well, exhibits a leveling. We certainly have not run sufficiently long to capture the very low frequency tail of the gravitational wave strain associated with prolate or oblate explosive outflows \cite{Murphy09,Yakunin10,Yakunin15}, though this is of secondary importance from a detection perspective. In this context, it is important to note that the fundamentally different generation mechanism of gravitational wave emission in our model -- i.e., excitation from deep within the proto-neutron star and not from accretion flows above the proto-neutron star surface -- should impact conclusions we might draw with regard to the post-bounce time evolved and whether or not the dominant gravitational wave emission has been captured.

Along with Powell and Mueller \cite{Powell2019}, we emphasize that we do not believe that the findings of the different groups \cite{Kuroda2016a,Andresen2017,O'Connor2018,Radice2019,Powell2019}, including the findings we present here, are necessarily 
in conflict but rather point to a potential model dependency and to a richer spectrum of core collapse supernova gravitational wave 
physics than could have been fully anticipated prior to the completion of these studies.

With regard to potentially SASI-associated gravitational wave emission, Torres-Forn\'{e} et al. \cite{Torres-Forne19} recently identified the low-frequency emission in their models with the fundamental $^{2}f$ mode, not with the SASI, and suspect that the low-frequency emission has been misclassified by others as SASI-induced. They further concluded that the fundamental mode is excited during periods of significant SASI activity and that its characteristics match perfectly with the time evolution of the shock -- i.e., the shock oscillates with the same frequency. They were puzzled by this, expecting instead that in the presence of the SASI the shock would oscillate with frequencies corresponding to the unstable modes of the vortical--acoustic cycle, not with a frequency corresponding to an acoustic cycle. Our results may shed some light on this discussion. Our low-frequency emission clearly stems from the convective gain layer and, therefore, must have its origins in either neutrino-driven convection or the SASI, or both. We have provided clear evidence of the SASI in our model and were able to correlate the SASI timescale we observe with the peak in the low-frequency gravitational wave spectrum. Guided by Torres-Forn\'{e} et al.'s analysis, we would naturally classify our low-frequency emission, with its origins in the convective layer, as p-mode emission -- i.e., as emission whose origin lies in acoustic modes. Thus, the SASI-induced emission in our model is p-mode emission.
Laming \cite{Laming2007} demonstrated that the SASI may result from {\em either} a vortical--acoustic cycle {\em or} a purely acoustic cycle, depending on the shock standoff radius. Foglizzo et al. \cite{Foglizzo2007} were able to conclude that a vortical--acoustic cycle is operative only for large standoff radii ($\geq10$). Laming came to the same conclusion for such large radii, but concluded that a purely acoustic mechanism, as discussed in Blondin and Mezzacappa \cite{Blondin06}, should operate at smaller standoff radii, such as those found in our model ($\leq3$). Looking at Figure 1 of Torres-Forn\'{e} et al. \cite{Torres-Forne19}, the largest standoff radius observed in either of their models -- s20 or 35OC -- over the course of the evolution of both models is $\sim$2.3, seen in model s20 between 100 and 200 ms after bounce, which, according to Laming, is in the range where an acoustic origin of the SASI should be expected.

Finally, our Ledoux unstable region -- i.e., our main source of gravitational radiation -- is deep within the proto-neutron star, far removed from the star's surface layers, and produces gravitational wave emission best characterized as p-mode emission. Thus, extracting information about the mass and radius of the proto-neutron star in our model from analytic estimates of surface g-mode \cite{Muller13} frequencies, whether the g-modes are excited from below \cite{Andresen2017} or from above \cite{O'Connor2018,Radice2019,Powell2019}, would not be appropriate. Further consideration on how to connect, if at all possible, the observed peak frequencies and important proto-neutron star parameters in our model and in others like it is warranted. Thus, our results suggest there is an even richer spectrum of possibilities, with gravitational radiation emission stemming from (i) surface g-modes excited by proto-neutron star convection, (ii) surface g-modes excited by flows in the gain layer, or (iii) p-modes within the proto-neutron star convective layer itself. In some cases, it may be more difficult, perhaps impossible, to extract proto-neutron star parameters, such as mass and radius. We also wish to acknowledge a caution raised by Powell and M\"{u}ller \cite{Powell2019}. We attribute our gravitational wave emission largely to Ledoux convection in the proto-neutron star. The direct correlation between the onset of Ledoux instability in the deep interior of the proto-neutron star and the onset of the high-frequency gravitational wave emission associated with it is obvious in our model. Regions of Ledoux instability will not give rise to other instabilities, such as the doubly-diffusive instabilities discussed by Bruenn and collaborators \cite{Bruenn1995,Bruenn1996}. Ambiguity does not arise in this sense. However, such instabilities cannot be accurately captured in our model given our use of ray-by-ray neutrino transport. This requires a three-dimensional treatment of neutrino transport. Thus, our model precludes an accurate development of such, more exotic modes and, in turn, any gravitational wave emissions associated with them. Future studies by us and by others should be conducted to exhaust the possibilities that may in fact become manifest in core collapse supernova environments.

Our gravitational wave analysis is based on data from our first three-dimensional core collapse supernova simulation, which is part of our \chimera C-series simulation suite. Thus, it is confined to a single progenitor mass and a single equation of state. Future work along these lines will be based on our follow-on D-series \chimera runs, which will span progenitor mass and metallicity (including the progenitor considered here) and which will be performed using a Yin--Yang angular grid with higher spatial resolution in both radius and angle. In particular, the 1-degree angular-resolution-equivalent, in both $\theta$ and $\phi$, in our D-series models, afforded by the use of a Yin--Yang grid, should be a notable improvement over the constant-$\mu$ grid employed here. Other equations of state will be considered, as well. We will report on the results of our gravitational wave analysis using the data from these models as they become available. This will allow us to validate the analysis presented in this work, as well as the conclusions drawn from it, while at the same time allowing us to assess quantitatively the impact of numerical resolution and other numerical issues on our gravitational wave emission predictions. It will also allow us to explore the variety of core collapse supernova gravitational wave emission and its underlying causes given different progenitors and input physics. 

The authors wish to thank the anonymous referee for a very careful reading of our manuscript and for questions that led to an improved analysis.The authors also wish to acknowledge fruitful discussions with Michele Zanolin, Marek Szczepanczyk, and Jasmine Gill.  This research was supported by the National Science Foundation Gravitational Physics Theory Program (PHY 1505933 and 1806692) and by the U.S. Department of Energy Offices of Nuclear Physics and Advanced Scientific Computing Research. This research was also supported by an award of computer time provided by the Innovative and Novel Computational Impact on Theory and Experiment (INCITE) Program at the Oak Ridge Leadership Computing Facility (OLCF) and at the Argonne Leadership Computing Facility (ALCF), which are DOE Office of Science User Facilities supported  under contracts DE-AC05-00OR22725 and DE-AC02-06CH11357, respectively. P. M. is supported by the National Science Foundation through its employee IR/D program. The opinions and conclusions expressed herein are those of the authors and do not represent the National Science Foundation. 

%

%

\end{document}